# Trigonometric parallaxes of star forming regions in the Sagittarius spiral arm

Y. W. Wu[1,2,3], M. Sato[2], M. J. Reid[4], L. Moscadelli[5], B. Zhang[2], Y. Xu[1], A. Brunthaler[2], K. M. Menten[2], T. M. Dame[4], X.W. Zheng[6],

[1] Purple Mountain Observatory, Chinese Academy of Sciences, Nanjing 210008, China
[2] Max-Planck-Institut für Radioastronomie, Auf demHügel 69, 53121 Bonn, Germany
[3] Graduate University of Chinese Academy of Sciences, Beijing 100049, China
[4] Harvard-Smithsonian Center for Astrophysics, 60 Garden Street, Cambridge, MA 02138, USA
[5] INAF, Osservatorio Astrofisico di Arcetri, Largo E. Fermi 5, 50125 Firenze, Italy
[6] School of Astronomy and Space Sciences of Nanjing University, Nanjing 210093, China

April 18, 2014

**ABSTRACT**

We report measurements of parallaxes and proper motions of ten high-mass star-forming regions in the Sagittarius spiral arm of the Milky Way as part of the BeSSeL Survey with the VLBA. Combining these results with eight others from the literature, we investigated the structure and kinematics of the arm between Galactocentric azimuth $\beta \approx -2°$ and $65°$. We found that the spiral pitch angle is $7°\!\!.3 \pm 1°\!\!.5$; the arm's half-width, defined as the rms deviation from the fitted spiral, is $\approx 0.2$ kpc; and the nearest portion of the Sagittarius arm is $1.4 \pm 0.2$ kpc from the Sun. Unlike for adjacent spiral arms, we found no evidence for significant peculiar motions of sources in the Sagittarius arm opposite to Galactic rotation.

Use \titlerunning to supply a shorter title and/or \authorrunning to suply a shorter list of author.

## 1. Introduction

Water ($H_2O$) masers at 22 GHz and class II methanol ($CH_3OH$) masers at 6.7 GHz and 12.2 GHz are found toward regions of massive star formation throughout the Milky Way. The brightest masers can reach flux densities of thousands of Jansky (Cheung et al. 1969; Menten 1991). Water masers are collisionally excited and are usually associated with shocked gas in jets and outflows launched by young stellar objects (YSOs), as well as from evolved stars (Hachisuka et al. 2006; Deacon et al. 2007). Class II methanol masers are produced by radiative pumping in warm, dense, and dusty environments around massive YSOs (Sobolev et al. 1997; Cragg et al. 2001, 2005; Goddi et al. 2011; Matsumoto et al. 2011).

Both water and methanol masers are compact enough to serve as excellent sources for astrometric observations. Recently, it has become feasible to measure trigonometric parallaxes of compact radio sources with accuracies of ∼ 10 $\mu$as with Very Long Baseline Interferometric (VLBI) networks like the NRAO[1] Very Long Baseline Array (VLBA) in the U.S.A., the VLBI Exploration of Radio Astrometry (VERA) array in Japan, and the European VLBI Network (EVN) in Europe and China (Xu et al. 2006; Honma et al. 2007; Rygl et al. 2010).

In addition to accurate distances and positions, multi-epoch VLBI phase-reference observations also provide proper motions, which, together with Doppler velocities, yield full three-dimensional space motions. This offers the possibility to reveal the structure and dynamics of the Milky Way disk in detail, as shown by Reid et al. (2009b) and Honma et al. (2012), who used "gold standard" distances and full phase-space information. Motivated by this progress, we started a large project in 2010 – the Bar and Spiral Structure Legacy (BeSSeL) [2] Survey, an NRAO key science project – with the aim of measuring accurate distances and proper motions for hundreds of high-mass star-forming regions (HMSFRs) across the Milky Way.

The Sagittarius spiral arm is the nearest arm inward from the Sun toward the Galactic center. This arm was first identified from blue giants by Morgan et al. (1953) and has since been seen in nearly all tracers of spiral structure such as H II regions (Georgelin & Georgelin 1976), H I (Henderson 1977) and CO (Dame et al. 2001) clouds, OB stars (Bronfman et al. 2000), far-IR emission from warm dust (Drimmel 2000), the Cepheids (Majaess et al. 2009), H$_\alpha$ (Lockman 1979), and [C II] and [N II] emission (Steiman-Cameron et al. 2010). In this paper, we report trigonometric parallax measurements of ten maser sources in HMSFRs in the Sagittarius arm. Combined with eight sources from the literature (see Table 2), we now have 18 masers with trigonometric parallaxes located in the Sagittarius arm.

The plan of this paper is as follow. In Section 2 we describe the observations and data analysis. Parallaxes and proper motions are presented in Section 3. The assignment of these sources to the Sagittarius arm is discussed in Section 4, where the geometry and kinematics of the Sagittarius arm are discussed. Conclusions are drawn in Section 5.

## 2. Observations and data analysis

We conducted parallax observations toward three 12 GHz methanol and eight 22 GHz water masers between 2010 and 2012 under VLBA programs BR134A and BR145B, C, L, J, N, T and W. Towards G037.43+01.51 both maser species were observed. We employed a phase-referencing tech-nique to measure the position offsets of the target masers relative to one or more background continuum sources nearby on the sky. The background sources were obtained from a calibrator search based on VLBA observations of compact NVSS (Condon et al. 1998) and CORNISH (Purcell et al. 2008) sources near our target masers (Immer et al. 2011). For 8 out of the 11 masers, we selected four background calibrators; for G043.79−00.12 three background calibrators were observed, and for G045.07+00.13 and G351.44+00.65 two background calibrators were observed (see Table 1).

---

[1] The National Radio Astronomy Observatory is a facility of the National Science Foundation operated under cooperative agreement by Associated Universities, Inc.
[2] www.mpifr-bonn.mpg.de/staff/abrunthaler/BeSSeL/index.shtml

Since most of our targets are water masers with individual spot lifetimes typically of a year or shorter, the observation dates were optimized for masers detectable for only seven months. Most masers were observed at six epochs during a period of one year. For G045.07+00.13, the observations were designed to simultaneously observe W 49 and the microquasar GRS 1915+105, and we observed at 16 epochs during 1.5 years. The dates of the observations are listed in Table A.1. The observational setup we used here is similar to that of Reid et al. (2009a). We employed four adjacent intermediate-frequency (IF) bands in dual circular polarization and set the maser's peak LSR velocity at the band center of the second IF. Each band was 8 MHz wide, and the band containing the maser emission was correlated to produce 256 spectral channels. The total bandwidth was wide enough to detect continuum sources (at the level of a few mJy beam$^{-1}$), and spectral channels of 31.25 kHz provided velocity spacings of 0.42 km s$^{-1}$ and 0.77 km s$^{-1}$ for the 22 GHz $H_2O$ and 12 GHz methanol masers, respectively.

The data correlation was performed with the DiFX[3] software correlator (Deller et al. 2007) at the array operations facility in Socorro, NM. The visibilities from each antenna pair were cross-correlated with an integration time of 0.9 s, which limited the instantaneous field of view of the interferometer with high sensitivity to a radius of ≈ 3" and 5" for the 22 and 12 GHz observations, respectively. The data reduction was conducted using the NRAO Astronomical Image Processing System (AIPS), together with scripts written in a ParselTongue-Python interface to AIPS and Obit (Kettenis et al. 2006). Calibration followed a four-step procedure described in Reid et al. (2009a). After calibration, we imaged the continuum emission of the background sources and target masers using the AIPS task IMAGR. Then we fitted elliptical Gaussian brightness distributions to images of background sources and cubes of maser emissions using the AIPS task JMFIT or SAD. In Figure A.1, we present the spectra of the target masers produced by averaging the cross-power spectra from the first-epoch data.

## 3. Results

### 3.1. Parallax and proper motions

In this section we present the parallax and proper motions of 11 maser sources toward ten HMSFRs. After we identified a maser spot during four or more epochs, we used a parallax sinusoid and a proper motion in each coordinate to model its position relative to the background sources. Since systematic errors due to unmodeled atmospheric delay variations, maser blending, and potential structure of background sources are usually larger than the formal position uncertainties, we added error floors to the east and north position offsets. When fitting, the error floors were adjusted until the residuals of parallax and proper motion fit gave a reduced $\chi^2_\nu$ of unity.

---

[3] DiFX, a software Correlator for VLBI, is developed as part of the Australian Major National Research Facilities Programme by the Swinburne University of Technology and operated under licence.

Usually, the data from several maser spots and background sources can be used to derive a parallax. First, we modeled position offsets of each maser spot relative to each background source to yield independent estimates. We discarded spots that yielded negative parallaxes or had parallax uncertainties > 0.050 mas. These usually came from complex, blended, maser spots. Then we removed more subtle outliers if they fell outside a range defined by $\overline{\Pi} \pm 2\sigma_\Pi$, where $\overline{\Pi}$ and $\sigma_\Pi$ are the average and standard deviation of parallaxes, respectively. Since one expects the same parallax for all maser spots, we combined the remaining spots to obtain a best estimate of the parallax of the source and the proper motion of each spot. The quoted parallax uncertainty is the formal error multiplied by $\sqrt{N}$, where $N$ is the number of maser spots fitted, to allow for the possibility of correlated position errors for the maser spots relative to the extragalactic continuum sources. The position data and the parallax and proper motion fits are shown in Figures 1–2.

The parallax residuals of G045.07+00.13 referenced to J1913+1220 were not randomly distributed around zero, but showed a small trend consistent with a position acceleration term. This was also seen in the microquasar GRS 1915+105 data, which was obtained simultaneously with G045.07+00.13 and used the same background source. When referenced to a different background source, J1913+0932, this acceleration term disappears from both G045.07+00.13 and GRS 1915+105. This suggests QSO J1913+1220 has slow structural changes that cause an apparent acceleration in the parallax data. Therefore, for G045.07+00.13, we only used J1913+0932 when estimating parallax and proper motion. We also checked for apparent accelerations, caused by time-variable background source structure, by plotting the difference between positions of background sources over time and found no other problems.

Methanol maser motions relative to a central exciting star are usually slow, $\sim$ 5 km s$^{-1}$ (Goddi et al. 2011), but for water masers they can be 10 and up to $\approx$ 100 km s$^{-1}$ for the most powerful sources (Hachisuka et al. 2006). Therefore, the fitted absolute proper motion of a given maser spot may deviate significantly from that of the central star for water masers. To estimate the influence of internal maser motions on the estimate of the central star's motion, we measured the proper motions (relative to the reference maser spot) of all spots detected in more than three epochs.

The proper motion of the central, exciting star was estimated as $\mu_{star} = \mu_{ref} + \mu_{off}$, where $\mu_{ref}$ is the reference spot's absolute proper motion and $\mu_{off}$ is the motion of the reference spot relative to the central star, estimated from the ensemble of maser spot motions as described below. The uncertainty of central star's proper motion is $\sigma_\mu = \sqrt{\sigma^2_{ref} + \sigma^2_{off}}$, where $\sigma_{ref}$ and $\sigma_{off}$ are uncertainties of $\mu_{ref}$ and $\mu_{off}$. For the H$_2$O maser G049.19−00.34, whose spots well sample an outflow, we fitted a radial expanding outflow model to solve for $\mu_{off}$ and $\sigma_{off}$. The model fitting results are given in Table A.3. For the H$_2$O maser sources G014.63−00.57, G035.02+00.34, G043.89−00.78 and G045.45+00.05 with few spots, and for G043.79−00.12 and G045.07+00.13, whose spot distribution is suggestive of multiple exciting stars, we used the uniformly weighted average of the relative proper motions of all spots as an estimate of $\mu_{off}$. For these $H_2O$ sources that could not be fitted with a simple expanding model, differences between $V_{LSR}$ of the water masers and CO lines from their parent molecular clouds were smaller than 10 km s$^{-1}$ (see discussion in Section 3.3), and we adopted a 10 km s$^{-1}$ uncertainty translated into a proper motion component uncertainty $\sigma_{off}$ = 10/D(kpc)/4.74 mas yr$^{-1}$.

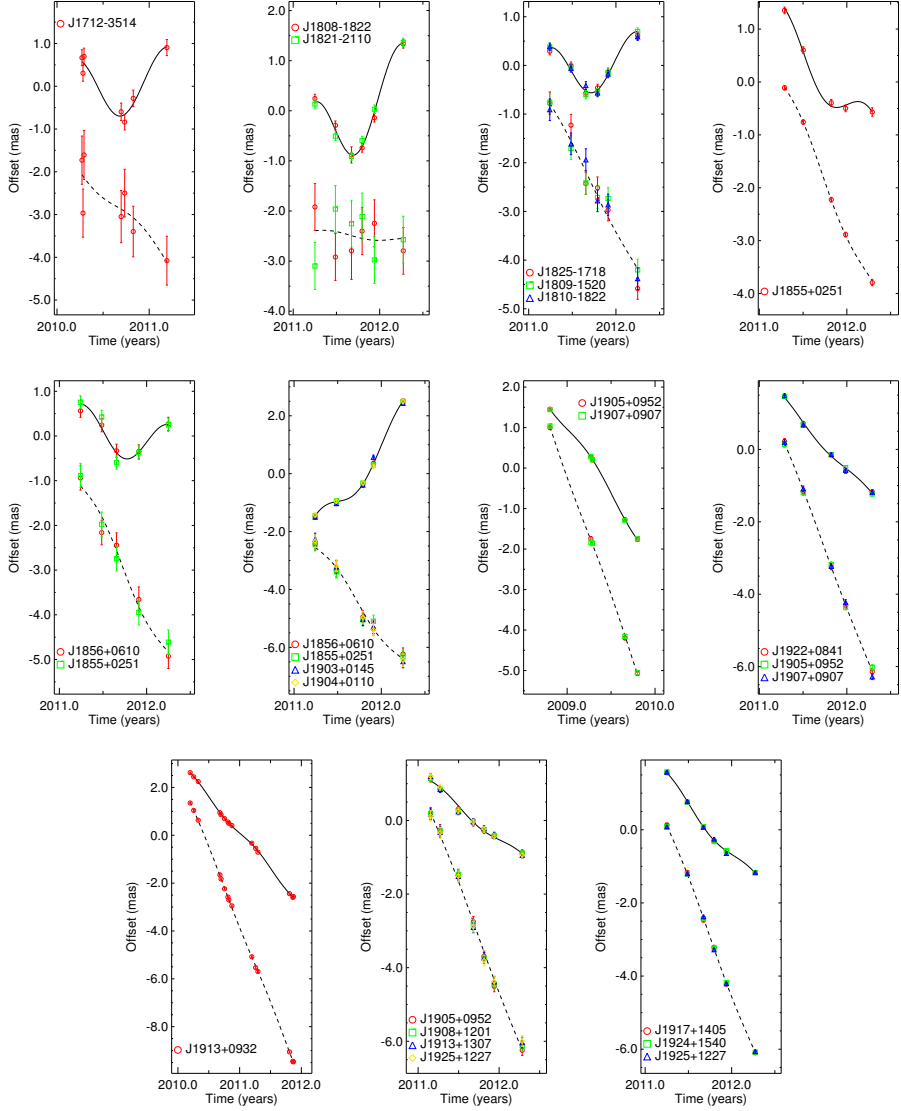

Fig. 1: Parallax and proper motion data and fits. Solid and dashed lines show the best-fitting parallax and proper motion model for eastward and northward offsets, respectively. *Top panels*: G351.44+00.65, G011.49−01.48, G014.63−00.57, and 22 GHz G035.02+00.34; *middle panels*: 12 GHz G037.43+01.51, 22 GHz G037.43+01.51, G043.79−01.51, and G043.89−00.78 *bottom panels*: G045.07+00.13, G045.45+00.05 and G049.19-00.34.

For the $CH_3OH$ maser source G351.44+00.65, we detected ten maser spots; the distribution were difficult to model by simple expansion, infall, or rotation. For the $CH_3OH$ masers G011.49−01.48 and G037.43+01.51, we only detected two and one maser spots, respectively. Since $CH_3OH$ masers generally move slowly (< 5 km s$^{-1}$) with respect to their central exciting star, we used the unweighted average of the relative proper motions of all spots for $\mu_{off}$. For these masers, the differences between $V_{LSR}$ of the methanol masers and CO lines are less than 5 km s$^{-1}$ (see discussion in Section 3.3); therefore, we adopted $\sigma_{off} = 5/D(kpc)/4.74$ mas yr$^{-1}$.

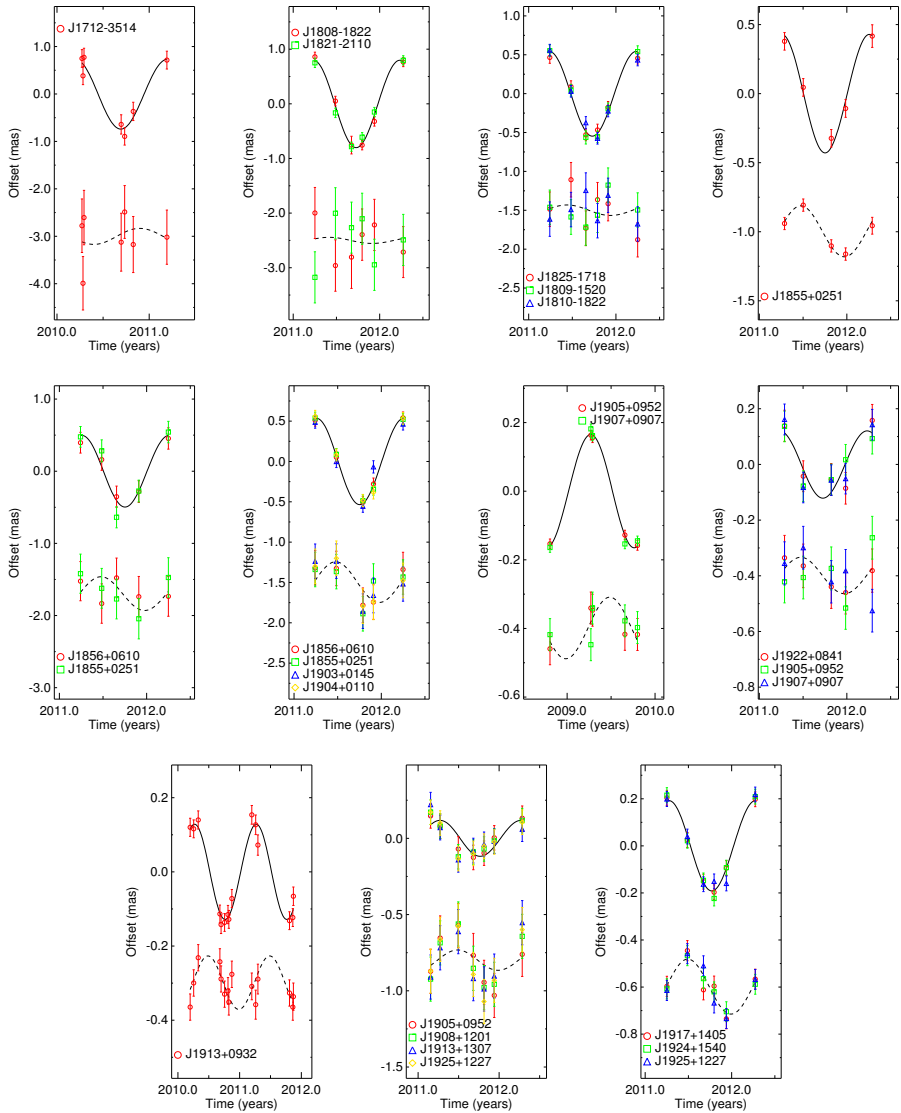

Fig. 2: Same as Fig. 1, but without the best-fit proper motion. *Top panels*: G351.44+00.65, G011.49−01.48, G014.63−00.57, and 22 GHz G035.02+00.34; *middle panels*: 12 GHz G037.43+01.51, 22 GHz G037.43+01.51, G043.79−01.51, and G043.89−00.78 *bottom panels*: G045.07+00.13, G045.45+00.05, and G049.19-00.34.

More details of the measurement of parallaxes and internal maser motions are presented in the online material (Tables A.2 and A.4).

3.2. Galactic Locations of Individual HMSFRs

Table 2 lists the parallax, proper motions, and LSR velocities of the 18 masers that we located in the Sagittarius arm, ten from this paper and eight from the literature. Some of the HMSFRs traced by masers may be associated with the same giant molecular cloud. For example, toward $\ell \approx 49°$ four maser parallaxes were measured: the 12 GHz methanol maser in W51 IRS2 and the H$_2$O masers in W51 main, G049.19−00.34 and G052.10+01.04. The first three of these sources are projected within the W51 star-forming complex and their distances are consistent. Their variance-weighted average parallax is 0.189±0.007 mas, corresponding to a distance of $5.29^{+0.19}_{-0.18}$ kpc. The remaining source, G052.10+01.04, is 1.3 kpc nearer than W 51 and has a 15 km s$^{-1}$ lower V$_{LSR}$. This means that it is very likely associated with a different giant molecular cloud a few degrees along the Sagittarius arm.

At $\ell \approx 15°$, parallaxes for three maser sources were measured: G014.33−00.64, G014.63−00.57 and G015.03−00.67. The distances of G014.63−00.57 ($1.83^{+0.08}_{-0.07}$ kpc) and G015.03−00.67 ($1.98^{+0.14}_{-0.12}$ kpc) are consistent within their uncertainties and their V$_{LSR}$ values agree within 5 km s$^{-1}$), and their linear separation on the sky is only 14 pc, suggesting that these sources are located in the same star-forming complex. The variance-weighted average parallax of these two measurements is 0.533±0.018 mas, corresponding to a distance of $1.87^{+0.07}_{-0.06}$ kpc. The distance of G014.33−00.64, 1.12 kpc, is significantly different from that of G014.63−00.57 and G015.03−00.67, suggesting that it resides in a different giant molecular cloud.

Towards $\ell \approx 35°$, five maser parallaxes were measured. G035.02+00.34 and G035.19−00.74, which although have similar distances, show a large difference in V$_{LSR}$ of 22 km s$^{-1}$ and a separation on the sky of 50 pc, which suggests that they belong to different HMSFRs. The other three masers, G034.39+00.22, G035.20−01.73, and G037.43+01.51, have significantly different distances: G034.39+00.22 at 1.56 kpc is associated with the filamentary MSX dark cloud G034.43+00.24 (Egan et al. 1998); G035.20−01.73 at 3.27 kpc is associated with the star-forming complex W48A; and G037.43+01.51 at 1.88 kpc is associated with the pre-main-sequence star IRAS 18517+0437.

Near Galactic longitude 45°, four maser parallaxes were obtained: G043.79−00.12, G043.89−00.78, G045.07+00.13 and G045.45+00.05. The distances of G043.89−00.78 ($8.26^{+1.64}_{-1.17}$ kpc), G045.07+00.13 ($7.75^{+0.44}_{-0.40}$ kpc), and G045.45+00.05 ($8.40^{+1.40}_{-1.05}$ kpc) are consistent within the uncertainties. The projected distance between G045.07+00.13 and G045.45+00.05 is 50 pc, but their V$_{LSR}$ values agree within 5 km s$^{-1}$ and CO emission also shows connections between them, suggesting that they might belong to the same giant molecular cloud. The variance-weighted average parallax of these two measurements is 0.128±0.006 mas, corresponding to a distance of $7.81^{+0.38}_{-0.35}$ kpc. The projected linear distance between G043.89−00.78 and G045.07+00.13 is 190 pc and no connecting CO emission filament is found, indicating that these are independent HMSFRs. G043.79−00.12 with a significantly different distance and V$_{LSR}$ also belongs to another HMSFR.

Finally, the remaining two masers, G351.44+00.65 and G011.49−01.48, are well separated and must be in different HMSFRs.

In Figure A.2 we show the 18 maser positions overlaid on gas/dust emissions of these HMSFRs. The color images and contours are taken from $^{13}$CO Galactic Ring Survey (GRS, Jackson et al. 2006), the APEX Telescope Large Area Survey of the GALaxy (ATLASGAL, Schuller et al. 2009) and the Wide-field Infrared Survey Explorer (WISE, Wright et al. 2010). The agreement of the $V_{LSR}$ and position between the maser line and $^{13}$CO line emission indicates that these masers pinpoint HMSFRs that are thought to trace spiral arms.

### 3.3. Local standard of rest velocity

$V_{LSR}$ values are used to assign the star-forming regions to a spiral arm and to derive 3D motions. Here we describe how we estimated the $V_{LSR}$ values and their uncertainties for the HMSFRs listed in Table 2.

Class II methanol masers emission comes from dense molecular cores, several hundreds of AUs from the central, exciting stars ( see example in Goddi et al. 2011; Matsumoto et al. 2011). Their velocities typically agree with thermally emitting molecules (CO) within 3 km s$^{-1}$. Therefore, for 12 GHz methanol masers we chose the $V_{LSR}$ of maser lines, with an uncertainty of 3 km s$^{-1}$, as indicating the motion of the exciting star. One exception is G035.19−00.74, whose maser LSR velocity differs from CO LSR velocities by ≈7 km s$^{-1}$, and we increased its uncertainty to 7 km s$^{-1}$.

For most water masers, whose $V_{LSR}$ values agree with CO within 5 km s$^{-1}$, for example, G014.63−00.57, and we adopted the CO $V_{LSR}$, with an uncertainty of 5 km s$^{-1}$, as that of the exciting star. For H$_2$O masers whose $V_{LSR}$ can deviate by more than 5 km s$^{-1}$, we used the average of the $V_{LSR}$ of maser and CO lines as the velocity of the central star, with the difference as the uncertainty. For H$_2$O masers in W51 MAIN, whose $V_{LSR}$ constrained with the expanding model of Sato et al. (2010b), the uncertainty is estimated to be ±4 km s$^{-1}$.

## 4. Discussion

### 4.1. Sagittarius arm assignment

In Figure 3 we plot the locations of HMSFRs on CO longitude-LSR velocity diagram ($\ell$-v) from Dame et al. (2001). The broad shaded region shows the loci of the Sagittarius arm as identified in 21 cm emission by Burton & Shane (1970) and in CO by Cohen et al. (1980). For $\ell > 20°$, we adapted the $\ell$-v curve from Figure 1 of Burton & Shane (1970), by extrapolating the near side of the arm to $\ell = 0°, V_{LSR} = 0$ km s$^{-1}$. Because a spiral arm has a non-zero width (1$\sigma$ arm deviation ∼ 0.2 kpc, see Sect. 4.3) and possibly non-circular motions, we allowed for a ±5 km s$^{-1}$ velocity width for the $\ell$-v loci of the Sagittarius arm. Error bars denote the $V_{VLSR}$ errors for these HMSFRs.

Thirteen HMSFRs were assigned to the Sagittarius arm because their longitudes and velocities match that of CO emission from the arm. From Figure 3, it can be seen that nearly all of these 13 HMSFRs with $\ell > 15°$ are within or close to the $\ell$-v loci of the arm. We then checked that their distances were reasonable for a Sagittarius arm member (and clearly not members of the Local or Scutum arms). Within about 15° longitude of the Galactic center, the $\ell$-v loci of several spiral arms converge. We have five HMSFRs within this range (denoted by squares) whose distances are between ∼ 1 and 2 kpc (see Table 2); we assigned these to the Sagittarius arm based on their locations near other HMSFRs in the arm.

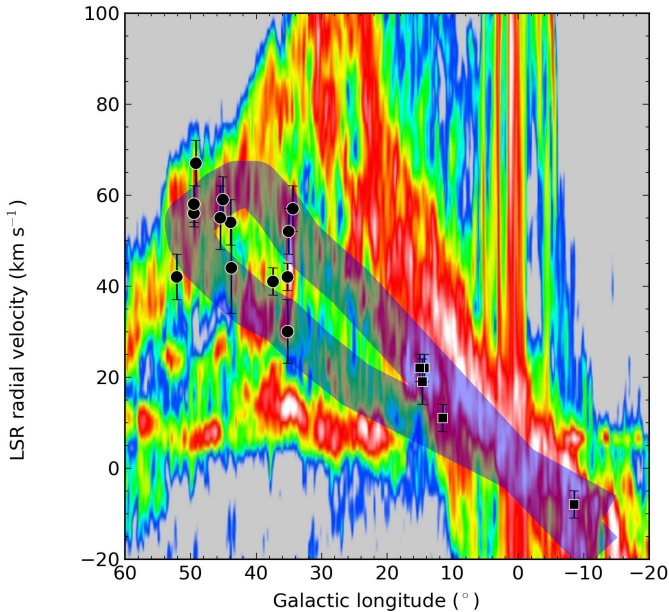

Fig. 3: Longitude-velocity diagram of CO emission from the CfA 1.2-m survey (Dame et al. 2001). The shaded region denotes the trajectory of the Sagittarius arm from Burton & Shane (1970). The width of the shaded region corresponds to a ±5 km s$^{-1}$ velocity dispersion. Dots indicate the HMSFRs assigned to the arm based on the longitude-velocity diagram; squares are HMSFRs assigned to the arm based on distances. Error bars reflect the V$_{LSR}$ uncertainties of the HMSFRs listed in Table 2.

*4.2. Geometry of the Sagittarius arm*

In Figure 4, we show locations of HMSFRs with measured trigonometric parallaxes on a plane view of the Galaxy. In addition to HMSFRs in the Sagittarius arm, we also plot HMSFRs in the Local arm (Xu et al. 2013) and the Scutum arm (Sato et al. 2013) for reference.

Because spiral arms wind around the center of the Galaxy, we transformed heliocentric distance, D, and Galactic longitude, $\ell$, to Galactocentric radius, R, and azimuth, $\beta$ (defined as 0 toward the Sun and increasing with Galactic longitude) to investigate the shapes of the arms. Values of $\beta$ and R for the observed 18 HMSFRs range from $-2°$ to $65°$ and 5.7 to 7.2 kpc (see Table 3), respectively. While the curvature of the Sagittarius arm was reported to vary smoothly in Georgelin & Georgelin (1976), the Taylor & Cordes model displays a flattening of the arm between $\beta = -20°$ and $65°$. We found no evidence for such a flattening from the maser parallaxes (see Figure 4). In the Taylor & Cordes spiral model, most distances were derived kinematically, which can result in errors of 30% or larger (Downes et al. 1980; Reid et al. 2009b), and these errors can be correlated across large portions of an arm. We suggest that the flattening of the Sagittarius arm in the Taylor & Cordes model may not be real but instead may be the result of fitting a cubic spline to inaccurate kinematic distances.

A spiral arm pitch angle, $\psi$, is defined as the angle between an arm segment and a tangent to a Galactocentric circle. For an ideal log-periodic spiral arm, $\ln(R) = \ln(R_{ref}) - (\beta - \beta_{ref}) \tan\psi$, where $\beta_{ref}$ is reference azimuth, arbitrarily set near the center of the distribution, and R$_{ref}$ is the radius of the arm at that azimuth. Because spiral arms may not be perfect geometrical structures, we initially used a Bayesian fitting procedure to estimate the pitch angle with an outlier-tolerant weighting scheme when calculating the probability distribution (see conservative formulation of Sivia & Skilling 2006), to estimate two parameters: $\ln(R_{ref})$ and $\psi$. Noting no significant outliers in these data, we then re-fitted with variance weighting (i.e., least-squares fitting). The (asymmetric) probability distribution function for the distance uncertainty was derived from a Gaussian distribution function for the measured parallaxes. In Figure 5, we show our best-fitting spiral model, where $\ln(R_{ref}) = 1.88 \pm 0.01$ at $\beta_{ref} = 25°.4$ and $\psi = 7°.3 \pm 1°.5$ (uncertainties give a 68% confidence range).

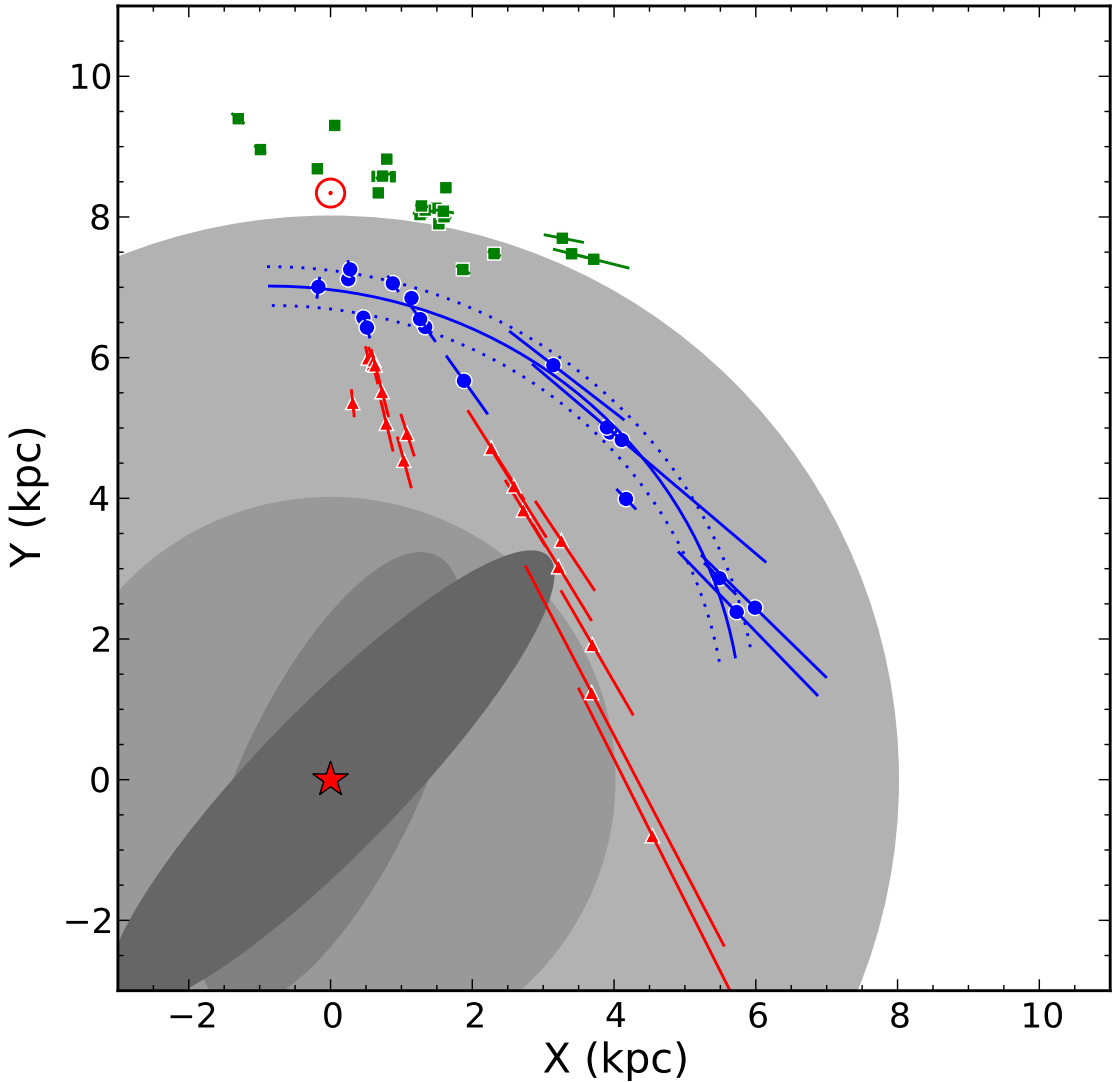

Fig. 4: Locations of HMSFRs (*blue dots*) determined by trigonometric parallaxes in the Sagittarius arm. The *green squares* and *red triangles* represent maser parallaxes in the Local arm (Xu et al. 2013) and Scutum arm (Sato et al. 2013). Error bars indicate $1\sigma$ distance uncertainties. The Galactic center is at (0,0) and the Sun at (0,8.33). Galactocentric azimuth, $\beta$, increases in clockwise direction, with $\beta = 0$ defined as the direction from the Galactic center toward the Sun. The background gray disks provide the scale, with radii corresponding to the Galactic bar region ($\approx 4$ kpc) and the solar circle ($\approx 8$ kpc). The short COBE boxy-bar and the long bar (Fux 1998; Blitz & Spergel 1991; Hammersley et al. 2000; Benjamin 2008) are indicated with shaded ellipses. The solid curved line traces the center (and the *dotted* lines the $\pm 1\sigma$ width) of the Sagittarius arm from the log-periodic spiral fitting (see Section 4.2). For this view of the Milky Way from the north Galactic pole, Galactic rotation is clockwise.

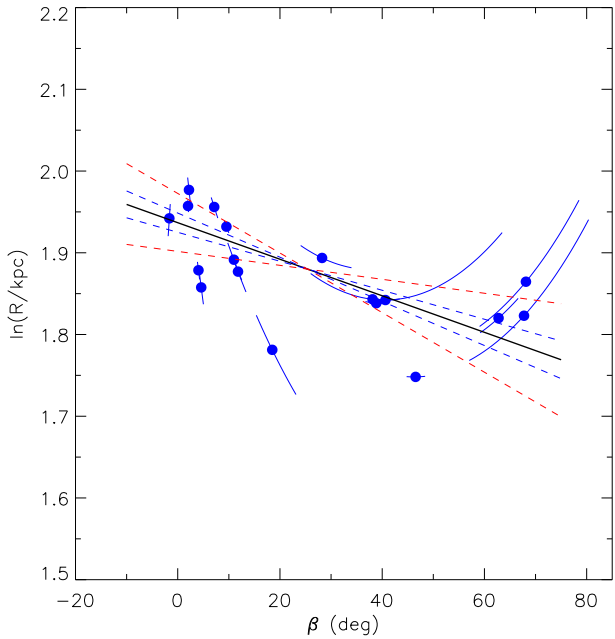

Fig. 5: Galactocentric radius, R, versus azimuth, $\beta$. The solid line shows the best-fitting model for a log-periodic spiral with a pitch angle $\psi_{Sag} = 7\overset{\circ}{.}3$; blue and red dash lines show the $\pm 1\sigma$ ($1\overset{\circ}{.}5$) and $\pm 3\sigma$ pitch angle uncertainties. Curved arcs denote the effects of $\pm 1\sigma$ parallax uncertainties.

These parameters place the tangent direction of the arm at a Galactic longitude of 51° and its nearest point to the Sun (toward $\ell \approx \beta \approx 0$) is 1.4 ± 0.2 kpc distant. When fitting the pitch angle, we adopted a Galactic center distance $R_0$ = 8.33 kpc (Gillessen et al. 2009; Reid et al. 2014). To assess the dependence of pitch angle on $R_0$, we repeated the fitting with $R_0$ = 8.0 kpc, to test whether variation on $R_0$ might cause large change of the pitch angle. Using $R_0$ = 8.0 kpc, we derived $\psi$ = $6\overset{\circ}{.}3 \pm 1\overset{\circ}{.}4$. Though slightly lower, this value is consistent with $7\overset{\circ}{.}3 \pm 1\overset{\circ}{.}5$.

Pitch angles based on maser parallaxes have been estimated for the Perseus arm, $\psi_{Perseus} = 9\overset{\circ}{.}5 \pm 1\overset{\circ}{.}3$ (Zhang et al. 2013); for the Outer arm, $\psi_{Outer} = 12°.1 \pm 4\overset{\circ}{.}2$ (Sanna et al. 2012); and for the Local arm, $\psi_{Local} = 10\overset{\circ}{.}1 \pm 2\overset{\circ}{.}7$ (Xu et al. 2013). Our pitch angle for the Sagittarius arm is marginally smaller than those of other arms measured recently by the BeSSeL Survey, and also smaller than the global average value of $12\overset{\circ}{.}8$ estimated by Vallée (2008). It is worth noting, however, that our measured pitch angle of $7\overset{\circ}{.}3 \pm 1\overset{\circ}{.}5$ agrees well with spiral fits to the arm in the first quadrant based on longitude-velocity diagrams of both H I ($6\overset{\circ}{.}5$; Burton & Shane 1970) and CO ($5\overset{\circ}{.}3$; Dame et al. 1986).

An arm width can be estimated from the full-width at half-maximum of the distribution of deviations of HMSFRs perpendicular to an arm model (see Column 7 in Table 3); the offsets range from −0.7 to 0.4 kpc, where a positive value locates a source radially outside the spiral arm model. In Figure 6, we show a histogram of these offsets. Except for G035.20−01.73, the other 17 HMSFRs are locating within ± 0.5 kpc of our best-fitting log-periodic spiral arm model. This is a rough estimate of the arm half-width at zero-probability.

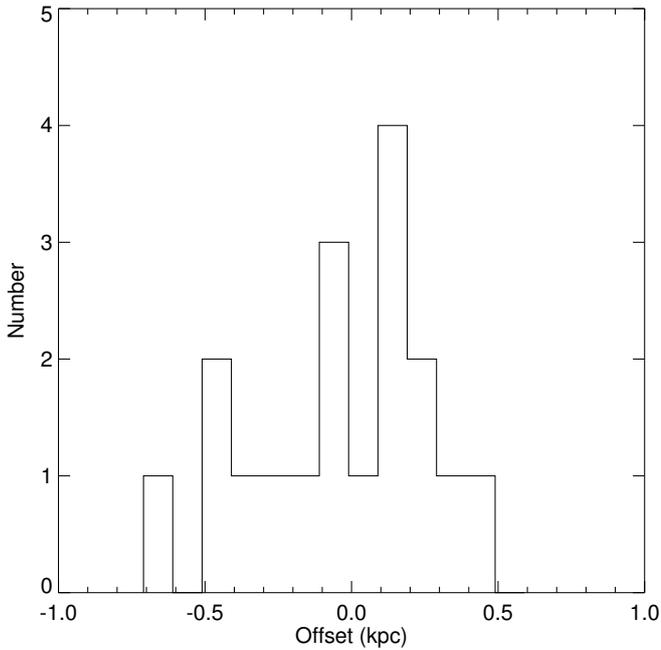

Fig. 6: Histogram of offsets of the HMSFRs perpendicular to the best-fitting log-periodic spiral arm model.

Since distance measurement uncertainties are typically about ±100 pc, they were expected to broaden the observed distribution somewhat. An alternative approach to estimating the width of the distribution of the arm tracers is to add the scatter expected from a non-zero arm width in quadrature to the measurement uncertainties, and then adjust that scatter term to achieve $\chi^2_\nu$ (per degree of freedom) of unity. We did this when fitting to the log-periodic spiral model and found we needed to add 0.21 kpc to the measurement uncertainty. This provides an estimate of the $1\sigma$ width of the arm that takes into account the contribution of the measurement uncertainty to the observed scatter.

### 4.3. 3D motions in the Sagittarius arm

From the source coordinates, distance (parallax), proper motion, and LSR velocity, we can determine both the 3-dimensional location and velocity vector for each source relative to the Sun. The velocity of the Sun relative to the Galactic center is the vector sum of a circular velocity of local standard of rest ($\Theta_0$) and the solar motion ($U_\odot$, $V_\odot$, $W_\odot$) relative to the local standard of rest. Given these parameters, one can then transform measured heliocentric motions to a Galactocentric reference frame. We can describe the space motions of the HMSFRs as a circular rotation component plus the non-circular (peculiar) velocity components, $U_s$, $V_s$, $W_s$, defined to be directed locally toward the Galactic center, in the direction of rotation, and toward the north Galactic pole, respectively (see the appendix in Reid et al. 2009b, for details).

In Table 4 we list the peculiar motion components of the HMSFRs, assuming a flat Galactic rotation curve, with Galactic constants $R_0$ = 8.33 ±0.16 kpc and $\Theta_0$ = 243 ± 6 km s$^{-1}$ and the solar motion values, $U_\odot = 10.7 \pm 1.8$ km s$^{-1}$, $V_\odot = 12.2 \pm 2.0$ km s$^{-1}$ and $W_\odot = 8.7 \pm 0.9$ km s$^{-1}$. Values of Galactic constants and solar motion are from the B1 fit of Reid et al. (2014), which were estimated by modeling parallaxes and proper motions of 80 HMSFRs with an axially symmetric model and with a tight prior for the solar motion parameters from Schönrich et al. (2010).

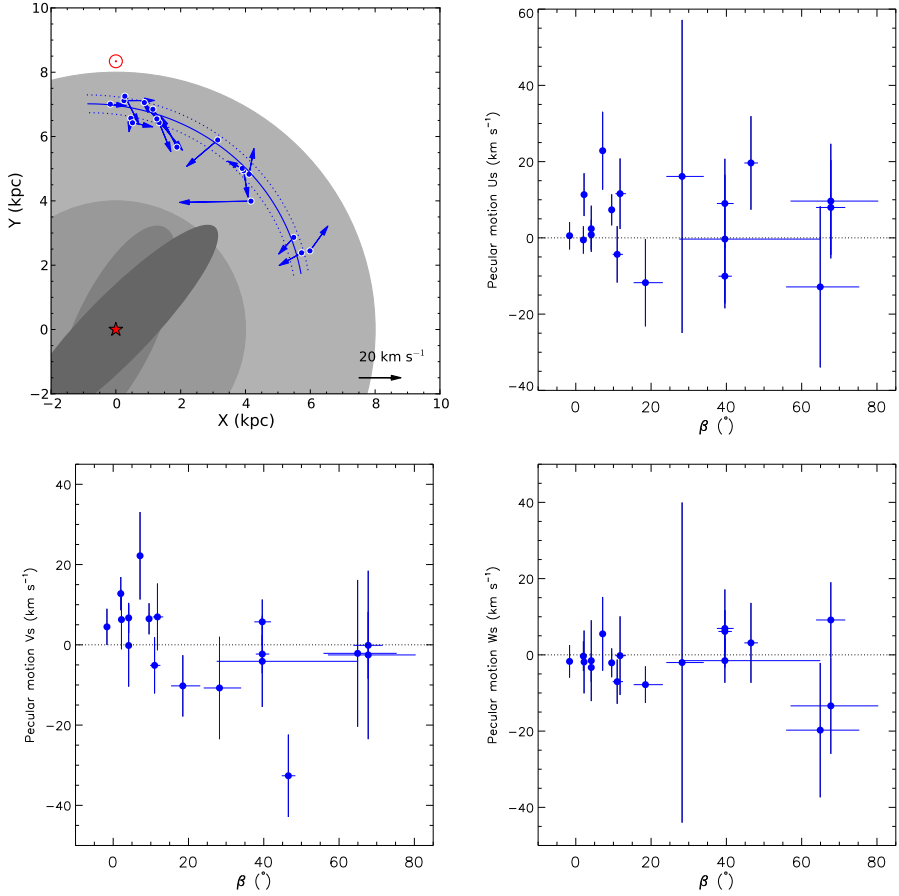

Fig. 7: Peculiar (non-circular) motions of HMSFRs. Upper left: vectors projected on the Galactic plane; the scale for motions is shown in the lower right corner of the panel. Upper right, lower left and lower right panels show peculiar motion components Us (toward the Galactic center), Vs (in the direction of Galactic rotation) and Ws (toward the north Galactic pole) versus Galactocentric azimuth ($\beta$). Peculiar motions assume $R_0 = 8.33 \pm 0.16$ kpc, $\Theta_0 = 243 \pm 6$ km s$^{-1}$, and solar motion components $(U_\odot, V_\odot, W_\odot) = (10.7 \pm 1.8, 12.2 \pm 2.0, 8.7 \pm 0.9)$ km s$^{-1}$.

The variance-weighted average peculiar motions for the 18 HMSFRs in the Sagittarius spiral arm are $\bar{U}_s = 2.9 \pm 1.5$ km s$^{-1}$, $\bar{V}_s = 3.7 \pm 1.5$ km s$^{-1}$, $\bar{W}_s = -1.7 \pm 1.4$ km s$^{-1}$. In Figure 7 we plot these peculiar motions as a function of Galactocentric azimuth. We do not find systematic trends in either the direction of rotation or toward the Galactic center.

We note that of the five parameters, $R_0$, $\Theta_0$, $U_\odot$, $V_\odot$ and $W_\odot$ used to derive peculiar motions, three are fairly well constrained (Gillessen et al. 2009; Brunthaler et al. 2011; Reid et al. 2014). However, there still is significant controversy about the values of the other two parameter: $\Theta_0$ and $V_\odot$ (Schönrich et al. 2010; Schönrich 2012; Brunthaler et al. 2011; Bovy et al. 2012; Reid et al. 2014). Honma et al. (2012) noted a strong correlation between the mean peculiar motion of HMSFRs, $\bar{V}_s$, and $V_\odot$. Were we to have adopted a higher value of $V_\odot$, e.g., $V_\odot = 20$ km s$^{-1}$, we would have found $\bar{V}_s = 10.8 \pm 1.5$ km s$^{-1}$ for the Sagittarius arm sources.

Although the peculiar motion component, $V_s$ depends critically on the value adopted for the solar motion in the direction of the Galactic rotation, $V_\odot$, one can compare peculiar motions among different spiral arms with little sensitivity to the assumed solar motion values. Re-calculating the peculiar motions of Local arm sources (Xu et al. 2013), using our adopted values listed above, we find $\bar{V}_s = -6.5 \pm 0.9$ km s$^{-1}$ for the Local arm. Using same Galactic parameters and solar motion values, Sato et al. (2013) estimated $\bar{V}_s = -8.3 \pm 2.0$ km s$^{-1}$ for the Scutum arm, and Choi et al. (2013) found $\bar{V}_s = -8.3 \pm 1.3$ km s$^{-1}$ for the Perseus arm. Thus, on average, the Sagittarius arm sources seem to be orbiting $\approx 10$ km s$^{-1}$ faster than sources in nearby arms.

In principle, different rotation curves might lead to different estimates of 3D motions for sources in different arms. If one assumes azimuthal symmetry (at least for Galactic radii > 4 kpc), as is implicitly done for a rotation curve, then there is fairly strong evidence for a relatively flat rotation curve for 5 < R < 15 kpc from maser data (see Reid et al. (2014)) and some other recent studies (e.g., Bovy & Rix (2013); Xin & Zheng (2013)). Also, Persic et al. (1996) argued fairly convincingly that essentially all spiral galaxies have similar forms for their rotation curves, which include a fairly flat region between about 0.5 to 2 $R_{opt}$, where $R_{opt}$ is the radius containing 83% of the optical light. This would translate to roughly 4 to 16 kpc for the Milky Way. Reid et al. (2014) explored in many forms for a rotation curve and only those with a relative constant rotation speed between Galactocentric radii of ∼ 5 and 16 kpc produced good fits. Therefore, we adopted a flat rotation curve, $\partial\Theta/\partial R = 0$ with $\Theta_0 = 243$ km s$^{-1}$.

Some published rotation curves are not flat. For example, McMillan (2011) investigated mass models for the Milky Way, using photometric and kinematic data. Their rotation curve peaks at R $\approx 8$ kpc. The decrease in rotation speed away from the peak is inconsistent with the maser parallax data summarized by Reid et al. (2014). If we use the McMillan best-fitting rotation curve (from their Fig. 5), the circular velocity at R = 6 kpc, a typical Galactocentric radius for the Sagittarius arm sources in this paper, the rotation speed is $\approx 7$ km s$^{-1}$ lower than at $R_0$. Decreasing $\Theta$ by this amount would not result in a significant change in the peculiar motion $V_s$ and would mostly affect the $U_s$. Thus even with a somewhat curved rotation curve, the conclusion that, on average, the Sagittarius arm sources seem to be orbiting $\approx 10$ km s$^{-1}$ faster than sources in nearby arms is still supported.

## 5. Conclusions

We have measured (trigonometric parallax) distances and proper motions for ten HMSFRs in the Sagittarius spiral arm. Complemented with eight more published results, these yield the location and shape of the arm, its pitch angle, and 3D motions of its components between Galactocentric azimuth $-2° < \beta < 65°$. We found the nearest portion of the Sagittarius arm is $1.4 \pm 0.2$ kpc from the Sun in the direction of the Galactic center. For the range of Galactocentric azimuth that we sampled, the arm has a pitch angle of $7°\!.3 \pm 1°\!.5$ and a $1\sigma$ width of $\approx 0.2$ kpc. We found no evidence of counter rotation, but did find that the Sagittarius arm sources orbit ∼ 10 km s$^{-1}$ faster than in adjacent arms.

While we now have a significant number of HMSFRs in the Sagittarius arm in the first Galactic quadrant, it would be interesting to extend these studies to more distant sources as well as to the adjoining fourth quadrant (where it may join the Carina arm). Such measurements are possible with a VLBI network in the southern hemisphere (such as the Australian Long Baseline Array and the forthcoming Square Kilometer Array).

This work was supported by the Chinese NSF through grants NSF 11133008, NSF 11073054, NSF 11203082, NSF 10921063, NSF 11233007, BK2012494, and the Key Laboratory for Radio Astronomy, CAS. This work was partially funded by the ERC Advanced Investigator Grant GLOSTAR (247078). We are grateful to James Urquhart for providing the ATLASGAL FITS files. We also acknowledge John D. Hunter, the creator of the Python *matplotlib*, which was used extensively for our figures.

*Facilities:* VLBA

**List of Objects**

## A.1. Appendix

Here we describe the star-forming regions whose parallaxes measured. In particular, we show their position in relation to compact dust emission imaged (mostly) at 870 $\mu$m from the APEX Telescope Large Area Survey of the GALaxy (ATLASGAL, Schuller et al. 2009). For the two sources at $b > 1.5°$, G035.20−01.73 and G037.43+01.51, which are not covered by ATLASGAL, we use 22 $\mu$m dust emission from the Wide-field Infrared Survey Explorer (WISE, Wright et al. 2010) as backgrounds. For HMSFRs that are covered by the $^{13}$CO Galactic Ring Survey (GRS, Jackson et al. 2006), we used GRS data as the background images, with ATLASGAL brightness shown as contours. In general, we found that the Class II CH$_3$OH masers are coincident with, or very close to, the peaks of the dust emission, consistent with the findings of Urquhart et al. (2013). The same is true for the H$_2$O masers, selected to be from HMSFRs.

### A.1.1. G351.44+00.65

G351.44+00.65 is associated with NGC6334I(N), a dense clump of the well-studied star-forming region complex NGC6334, also known as the Cat's Paw Nebula (Beuther et al. 2008; Persi & Tapia 2010). Its V$_{LSR}$ estimated from NH$_3$ thermal emission is −5 km s$^{-1}$ (Beuther et al. 2005).

We observed two background QSOs, but only the data from the one nearest to the maser target, J1712−3514, were used for the parallax fitting. Combining data from seven distinct 12 GHz maser spots, the best estimate of the parallax value is 0.744±0.076 mas. This corresponds to a distance of 1.35$_{-0.13}^{+0.15}$ kpc, 20% smaller than the photometric distance of 1.74 ± 0.31 kpc (Neckel 1978) and the near kinematic distance of 1.7 kpc (to derive the kinematic distances of the targets we employed the Galactic rotation model from Brunthaler et al. (2011), which uses the rotation parameters R$_0$ = 8.33 kpc and $\Theta_0$ = 243 km s$^{-1}$.). The measured distance places G351.44+00.65 at a Galactocentric radius of 6.97$_{-0.15}^{+0.12}$ kpc (using R$_0$ = 8.33 kpc).

We identified ten maser spots toward this source, but the distribution was difficult to model by simple expansion, infall, or rotation. This is expected for a methanol maser source. By adding the mean relative proper motion of (0.11 ± 0.55, −0.31 ± 0.55) mas yr$^{-1}$ (Table A.4) to the absolute proper motion of maser feature 1, (0.20 ± 0.19, −1.86 ± 0.71) mas yr$^{-1}$, we obtain an absolute proper motion of ($\bar{\mu}_X, \bar{\mu}_Y$) = (0.31 ± 0.58, −2.17 ± 0.90) mas yr$^{-1}$ for the exciting star (Table 2).

### A.1.2. G011.49−01.48

G011.49−01.48 is an active HMSFR containing the bright IRAS source 18134−1941 with a luminosity of ∼4 × 10$^3$L$_\odot$ and methanol maser emission. The V$_{LSR}$ is 11 km s$^{-1}$ (from NH$_3$ line, Sunada et al. 2007). The spectrum of the observed 12.2 GHz methanol masers consists of two features, with the bright one at 9.1 km s$^{-1}$ and the faint one at 15.2 km s$^{-1}$. The position of the maser agrees within 0″.5 with a 1.2 mm dust source (Hill et al. 2005).

The four background sources J1808−1822, J1821−2110, J1825−1817, and J1832−2039 were detected in all epochs. The derived parallax values from these four QSOs are consistent, ranging from 0.77 to 0.95 mas (Table A.2). Since the J1825−1817 and J1832−2039 have larger separations (> 3°) from the maser than the other two calibrators, the parallax uncertainties obtained with these two sources are about five times larger than those with the other two sources. Therefore, we only used the two nearby background sources J1808−1822 and J1821−2110 to estimate the parallax and derived a value of 0.800±0.033 mas, corresponding to a distance of $1.25^{+0.05}_{-0.05}$ kpc. In Figure 1 we showed the parallax and proper motion fits. Our distance is 30% smaller than the near kinematic distance of 1.6 kpc. The measured distance places G011.49−01.48 at a Galactocentric radius of $7.08^{+0.05}_{-0.05}$ kpc (using $R_0$ = 8.33 kpc).

Towards G011.49−01.48, we identified two methanol maser spots. At a distance of $1.25^{+0.05}_{-0.05}$ kpc, the feature separation is 346±4 AU and their relative velocity is 5.7 ± 0.5 km s$^{-1}$ (Fig A.3). The absolute proper motion of maser feature 1 is 1.16 ± 0.07 mas yr$^{-1}$ toward the east and −0.16 ± 0.41 mas yr$^{-1}$ toward the north. Adding the mean relative proper motion of (0.26±0.51, −0.45±0.51) mas yr$^{-1}$, we derived an absolute proper motion of $(\bar{\mu}_X, \bar{\mu}_Y)$ = (1.42 ± 0.52, −0.61 ± 0.65) mas yr$^{-1}$ for the exciting star (Table 2).

### A.1.3. G014.33−00.64

G014.33−00.64 is a Galactic star-forming region, first identified from the far-infrared 70-$\mu$m Galactic plane survey (Jaffe et al. 1982). $V_{LSR}$ of this region is about 22 km s$^{-1}$. Sato et al. (2010a) measured the water maser parallax and proper motion with VERA.

### A.1.4. G014.63−00.57

G014.63−00.57 is an HMSFR containing both 22 GHz water and class I methanol masers (Val'tts et al. 2000). The observed 22 GHz water maser presents a spectrum with only one peak with $V_{LSR}$ of 19 km s$^{-1}$. The maser position coincides with an extended green object (EGO), which has been classified as a massive young stellar object (MYSO) outflow candidate (Cyganowski et al. 2008).

Three (J1825−1817, J1809−1520, and J1810−1626) of the four background sources were detected in all the epochs. Parallax values determined with individual background sources were consistent, ranging from 0.49 to 0.61 mas. Figure 1 presents the parallax and proper motion fit of 0.546 ± 0.022 mas, combining the data from three background sources.

We identified six water maser spots (Table A.4). These spots trace a bipolar outflow with a size of 800 AU and speed of 30 km s$^{-1}$ (Fig A.3). The mean relative motion (with respect to feature 1), obtained by averaging the motion of the spots in the blue and red lobes separately and then averaging these mean values, is (−0.11±1.20, 1.31±1.20) mas yr$^{-1}$. By adding the mean relative proper motion to the absolute proper motion, (0.33±0.03, −3.38±0.09) mas yr$^{-1}$ of feature 1, we obtain an absolute proper motion of $(\bar{\mu}_X, \bar{\mu}_Y)$ = (0.22±1.20, −2.07±1.20) mas yr$^{-1}$ for the exciting star (Table 2).

*A.1.5. G015.03−00.67*

The CH$_3$OH maser toward G015.03−00.67 is from the giant molecular cloud M17 SW with a V$_{LSR}$ of about 22 km s$^{-1}$. Xu et al. (2011) measured the maser parallax and proper motion with the VLBA.

*A.1.6. G034.39+00.22*

The H$_2$O masers toward G034.39+00.22 are from an infrared dark cloud MSXDC G034.43+00.24, which is associated with the prototypical ultra-compact HII region G34.3+0.2 (Reid & Ho 1985). The V$_{LSR}$ is about 57 km s$^{-1}$. Kurayama et al. (2011) reported a maser parallax of 0.643± 0.049 mas as measured with VERA.

*A.1.7. G035.02+00.34*

G035.02+00.34 is an active HMSFR, containing hydroxyl, water, and methanol masers. The LSR velocity of the molecular cloud (from $^{13}$CO GRS survey) hosting this star-forming region is 52 km s$^{-1}$. The water masers are found in two regions of emission, with brighter spots near V$_{LSR}$ = 45 km s$^{-1}$ and fainter ones near V$_{LSR}$ = 56 km s$^{-1}$, perhaps tracing a bipolar outflow.

Of the four background sources, J1903+0145 and J1857−0048 were not detected in the last epoch; parallax uncertainties derived from J1904+0110 were larger than 0.05 mas, probably due to its large separation (> 2°) from the maser target, which resulted in large systematic errors. Thus, only data from J1855+0251 were used to fit the parallax and proper motion (Figure 1). The fits, combining the positions of the two maser spots, 1 and 6 in Table A.4, which were detected in all five epochs, and allowing for different proper motions, yielded a parallax of 0.431 ± 0.040 mas. Our trigonometric parallax for G035.02+00.34 corresponds to a distance of 2.32 $^{+0.24}_{-0.20}$ kpc, which places it at a Galactocentric radius of 6.54 $^{+0.14}_{-0.16}$ kpc (using R$_0$ = 8.33 kpc).

In total, we identified 17 water maser spots at different epochs. For 11 of them that were detected in at least three successive epochs, we derived relative proper motions. The distribution and internal motion of these spots indicate a bipolar outflow with a size of 650 AU and a speed of 25 km s$^{-1}$ (Fig A.3). By adding the mean relative proper motion of (−0.91 ± 0.90, −0.40 ± 0.90) mas yr$^{-1}$ (Table A.4) to the absolute proper motion, (−0.01 ± 0.10, −3.21 ± 0.07) mas yr$^{-1}$, of maser feature 1, we obtain an absolute proper motion of ($\bar{\mu}_X$, $\bar{\mu}_Y$) = (−0.92 ± 0.90, −3.61 ± 0.90) mas yr$^{-1}$ for the exciting star(Table 2).

*A.1.8. G035.19−00.74*

The CH$_3$OH maser G035.19−00.74 is associated with a bright IRAS source 18556+0136. The V$_{LSR}$ of this region is about 35 km s$^{-1}$. Zhang et al. (2009) measured the maser parallax and proper motion with the VLBA.

*A.1.9. G035.20−01.73*

The $CH_3OH$ maser G035.20−01.73 is associated with H II region W48A. The $V_{LSR}$ is about 43 km s$^{-1}$. Zhang et al. (2009) measured the maser parallax and proper motion with the VLBA.

*A.1.10. G037.43+01.51*

G037.43+01.51 contains a bright IRAS source 18517+0437 with a luminosity of ∼1 × 10$^4$ L$_\odot$, and water and Class II methanol masers. The $V_{LSR}$ is 44 km s$^{-1}$ (from $C^{18}O$ (2-1), López-Sepulcre et al. 2010). We observed both the 12.2 GHz methanol and the 22 GHz water masers. The positions and central velocities of these two masers agreed within 0.2 arcsec and 1 km s$^{-1}$, respectively.

At 22 GHz all four background sources were detected, but at 12 GHz we detected only two of them, J1856+0610 and J1855+0251. In Figure 1, we show the parallax and proper motion fits for methanol and water masers, respectively. The parallax results from these two maser species were consistent (Table A.2), although the methanol masers yielded larger uncertainties, which may be due to the larger image sizes owing to interstellar scattering. A combined fitting using both maser species, allowing for different proper motions of the water and methanol masers, yielded a parallax of 0.532±0.021 mas. This corresponds to a distance of 1.88 $^{+0.08}_{-0.07}$ kpc, which places G037.43+01.51 at a Galactocentric radius of 6.90 $^{+0.05}_{-0.05}$ kpc (using $R_0$ = 8.33 kpc).

The absolute proper motion of the methanol maser (usually within about ±5 km s$^{-1}$ of the exciting star) is (−0.45 ± 0.08, −3.69 ± 0.20) mas (see table A.4). Adding an uncertainty of 5 km s$^{-1}$ in transferring the maser to the central star's motion, yields a proper motion of (−0.45 ± 0.35, −3.69 ± 0.39) mas for the central star. (Table 2).

*A.1.11. G043.79−00.12*

G043.79−00.12, also known as OH 43.8−0.1, exhibits hydroxyl, water, and class II methanol masers. The $V_{LSR}$ is 44 km s$^{-1}$ (from $^{13}CO$ emission). The water maser has tens of emission features with $V_{LSR}$ between 30 to 60 km s$^{-1}$.

The three background sources were detected at all epochs. Position residuals of J1913+0932 showed a larger scatter than for the other two background sources, because this source was heavily resolved on long baselines. The best estimate of the parallax, combining the results for J1905+0952 and J1907+0907 is 0.166 ± 0.005 mas, corresponding to a distance of 6.02$^{+0.26}_{-0.24}$ kpc, which places G043.79−00.12 at a Galactocentric radius of 5.74$^{+0.004}_{-0.002}$ kpc (using $R_0$ = 8.33 kpc). The parallax and proper motion fitting is shown In Figure 1.

We identified 12 distinct water maser spots, distributed over ∼ 0.2″. The distribution and proper motions of these spots appear fairly random and we did not model an outflow (Fig A.3). By adding the mean relative proper motion, (0.21 ± 0.35, −0.06 ± 0.35) mas yr$^{-1}$, to the absolute proper motion, (−3.23 ± 0.01, −6.14 ± 0.04) mas yr$^{-1}$ of maser feature 1, we derive an absolute proper motion of $(\bar{\mu}_X, \bar{\mu}_Y) = (-3.02 \pm 0.36, -6.20 \pm 0.36)$ mas yr$^{-1}$ for G043.79−00.12 for the central star (Table 2).

### A.1.12. G043.89−00.78

G043.89−00.78 contains water and class II methanol masers (Szymczak et al. 2000; Hofner & Churchwell 1996). The V$_{LSR}$ is 54 km s$^{-1}$ (from the $^{13}$CO GRS survey). We detected water masers from 45 to 65 km s$^{-1}$.

Although all background sources were detected at all epochs, only the three sources nearest in angle to the masers (with a separation smaller than 3°) yielded consistent parallaxes. The best estimate of the parallax, combining the results from three background sources, is 0.121 ± 0.020 mas. This corresponds to a distance of $8.26^{+1.64}_{-1.17}$ kpc, which places G043.89−00.78 at the far side of the Sagittarius arm with a Galactocentric radius of $6.19^{+0.77}_{-0.33}$ kpc (using R$_0$ = 8.33 kpc).

We obtained proper motions for only four maser spots. By adding the mean relative proper motion of (−0.11 ± 0.30, −0.13 ± 0.30) mas yr$^{-1}$ (the uncertainty given here is a realistic uncertainty corresponding to a velocity of 10 km s$^{-1}$) to the absolute proper motion (−2.64 ± 0.04, −6.30 ± 0.06) mas yr$^{-1}$ of maser feature 3, we obtained an absolute proper motion of $(\bar{\mu}_X, \bar{\mu}_Y) = (-2.75 \pm 0.30, -6.43 \pm 0.30)$ mas yr$^{-1}$ for G043.89−00.78 for the central star (Table2).

### A.1.13. G045.07+00.13

G045.07+00.13 hosts numerous ultra-compact H II regions, maser sources, and outflows from young and highly embedded OB stellar clusters (Rivera-Ingraham et al. 2010). The V$_{LSR}$ is 59 km s$^{-1}$ (from the 13CO GRS survey). The water masers we observed consist of tens of emission features for 55 < V$_{LSR}$ < 65 km s$^{-1}$. For this source, we observed at 16 epochs spanning 1.5 years. The best estimate of the parallax is 0.129 ± 0.007 mas. This corresponds to a distance of $7.75^{+0.44}_{-0.40}$ kpc, which places it at the far side of the Sagittarius arm, with a Galactocentric radius of $6.17^{+0.15}_{-0.11}$ kpc (using R$_0$ = 8.33 kpc).

We measured proper motions for 12 spots. The distribution of these spots were difficult to model with a single outflow (Fig A.3). Therefore we averaged the motions to derive a mean relative proper motion, (−0.23 ± 0.26, 0.38 ± 0.26) mas yr$^{-1}$. Adding the absolute proper motion of (−2.98 ± 0.01, −6.49 ± 0.02) mas yr$^{-1}$ of maser feature 2, we derived an absolute proper motion of $(\bar{\mu}_X, \bar{\mu}_Y) = (-3.21 \pm 0.26, -6.11 \pm 0.26)$ mas yr$^{-1}$ for the central driving star (Table 2).

### A.1.14. G045.45+00.05

The water masers toward G045.45+00.05 may belong to the same molecular cloud complex as G045.07+00.13, with a projected separation of about 56 pc. The water masers have emission features with V$_{LSR}$ near 50 km s$^{-1}$, blueshifted relative to thermal line emission, with V$_{LSR}$ = 59 km s$^{-1}$ from the $^{13}$CO GRS survey.

We observed four background sources, all of which were detected at all five epochs. Parallax fits based on individual background sources ranged from 0.10 to 0.15 mas. The best estimate of the parallax, combining the four background sources, is 0.119 ± 0.017 mas. This corresponds to a distance of $8.40^{+1.40}_{-1.05}$ kpc, which is consistent with distance of G045.07+00.13 and supports their association.

We measured the proper motions of four maser spots. Averaging these, we obtained a mean relative proper motion of (−0.55 ± 0.38, 0.22 ± 0.54) mas yr$^{-1}$. Adding the absolute proper motion of (−1.79 ± 0.04, −5.78 ± 0.07) mas yr$^{-1}$ of maser feature 2, we derived an absolute proper motion of $(\bar{\mu}_X, \bar{\mu}_Y)$ = (−2.34 ± 0.38, −6.00 ± 0.54) mas yr$^{-1}$ for the central driving star (Table 2).

### A.1.15. W 51 region

The W51 star-forming region is near the tangent point of the Sagittarius arm. Toward this region are three maser sites with parallax measurements: the methanol maser G049.48−00.36 toward W51 IRS2 that was measured with the VLBA by Xu et al. (2009), the water maser G049.49−00.38 toward W51 main/south that was measured with the VLBA by Sato et al. (2010b), and the water maser G049.19−00.34 reported in this paper.

G049.19−00.34 belongs to the high-velocity stream in the W51 star-forming complex (Carpenter & Sanders 1998; Parsons et al. 2012). The position of the maser emission agrees with the millimeter dust continuum source BGPS 6312. The $V_{LSR}$ is 68 km s$^{-1}$ (from the $^{13}$CO GRS survey). The spectrum of the maser emission displays four features with 55 < $V_{LSR}$ < 85 km s$^{-1}$.

The four background sources were detected at all the epochs. The parallaxes derived individually from the four background sources are consistent, ranging from 0.170 to 0.227 mas. Probably owing to a large separation (> 2°) from the target maser, the parallax determined using J1930+1532 as a phase reference shows higher residuals than that of the other three background sources. Our best estimate of the parallax, combining data from the three background sources nearest the maser, is 0.192±0.009 mas. This corresponds to a distance of $5.21^{+0.25}_{-0.23}$ kpc, consistent with the recently derived parallax distance to W51 main/south of $5.41^{+0.31}_{-0.28}$ kpc (Sato et al. 2010b) and the distance to W51 IRS2 $5.12^{+2.94}_{-1.37}$ kpc (Xu et al. 2009).

Toward G049.19−00.34, we identified 16 water maser spots. The distribution and proper motions of these spots suggest a bipolar outflow, with a size of 290 AU and an outflow speed of 40 km s$^{-1}$ (Fig A.3). We fitted their proper motions with an expansion model, with results given in Table A.3. By adding the fitted relative proper motion of the central star to the absolute proper motion of maser feature 1, we derived an absolute proper motion of $(\bar{\mu}_X, \bar{\mu}_Y)$ = (−3.06 ± 0.40, −5.69 ± 0.40) mas yr$^{-1}$ for the central star in G049.19−00.34.

*A.1.16. G052.10+01.04*

The $H_2O$ maser toward G052.10+01.04 is associated with the bright IRAS source 19213+1723. The $V_{LSR}$ of this region is about 42 km s$^{-1}$. Oh et al. (2010) measured the maser's parallax and proper motion with VERA.

## A.1. ONLINE MATERIAL

Table 1: Source information

| Source | R.A.(J2000) (h m s) | Dec.(J2000) (°,′,″) | $\theta_{sep}$ (°) | P.A. (°) | Beam Size (mas × mas at °) | $F_{peak}$ (Jy beam$^{-1}$) | $V_{peak}$ (km s$^{-1}$) |
|---|---|---|---|---|---|---|---|
| G351.44+00.65 | 17 20 54.6080 | −35 45 08.612 | ... | ... | 7.33 × 1.99 at 15 | 12.89 | −10.0 |
| J1712−3514 | 17 12 05.1353 | −35 14 34.183 | 1.9 | −74 | 7.46 × 2.32 at 17 | 0.039 | ... |
| J1733−3722 | 17 33 15.1930 | −37 22 32.396 | 3.0 | 123 | ... | ... | ... |
| G011.49−01.48 | 18 16 22.1327 | −19 41 27.219 | ... | ... | 2.95 × 0.96 at −5 | 1.536 | 9.1 |
| J1808−1822 | 18 08 55.5154 | −18 22 53.396 | 2.2 | −53 | 5.21 × 1.87 at −18 | 0.031 | ... |
| J1821−2110 | 18 21 05.4692 | −21 10 45.262 | 1.9 | 143 | 4.83 × 1.74 at −19 | 0.055 | ... |
| J1825−1718 | 18 25 36.5323 | −17 18 49.848 | 3.2 | 42 | 5.01 × 1.58 at −11 | 0.155 | ... |
| J1832−2039 | 18 32 11.0465 | −20 39 48.203 | 3.8 | 105 | 5.11 × 1.58 at −8 | 0.233 | ... |
| G014.63−00.57 | 18 19 15.5407 | −16 29 45.786 | ... | ... | 0.94 × 0.30 at −10 | 3.757 | 19.4 |
| J1825−1718 | 18 25 36.5323 | −17 18 49.848 | 1.7 | 118 | 1.85 × 0.57 at −5 | 0.068 | ... |
| J1809−1520 | 18 09 10.2100 | −15 20 09.700 | 2.7 | −64 | 1.81 × 0.59 at −3 | 0.020 | ... |
| J1810−1626 | 18 10 39.8500 | −16 26 52.900 | 2.0 | −88 | 1.89 × 0.63 at −3 | 0.046 | ... |
| J1808−1822 | 18 08 55.5200 | −18 22 53.400 | 3.1 | 127 | 1.98 × 0.64 at −3 | 0.013 | ... |
| G035.02+00.34 | 18 54 00.6576 | +02 01 19.217 | ... | ... | 0.87 × 0.43 at −16 | 4.499 | 56.5 |
| J1855+0251 | 18 55 35.4364 | +02 51 19.563 | 0.9 | 25 | 0.98 × 0.33 at −15 | 0.097 | ... |
| J1903+0145 | 19 03 53.0633 | +01 45 26.311 | 2.5 | 96 | 1.05 × 0.34 at −14 | 0.024 | ... |
| J1857−0048 | 18 57 51.3586 | −00 48 21.950 | 3.0 | 161 | 1.12 × 0.55 at −15 | 0.011 | ... |
| J1904+0110 | 19 04 26.3973 | +01 10 36.709 | 2.7 | 108 | 1.06 × 0.34 at −13 | 0.036 | ... |
| G037.43+01.51[a] | 18 54 14.3481 | +04 41 39.647 | ... | ... | 1.97 × 1.15 at 11 | 0.946 | 41.0 |
| J1856+0610 | 18 56 31.8388 | +06 10 16.765 | 1.6 | 21 | 3.90 × 2.20 at 67 | 0.096 | ... |
| J1855+0251 | 18 55 35.4364 | +02 51 19.563 | 1.9 | 169 | 3.96 × 2.17 at 65 | 0.024 | ... |
| J1903+0145 | 19 03 53.0633 | +01 45 26.311 | 3.8 | 140 | 3.85 × 2.92 at 64 | 0.011 | ... |
| J1904+0110 | 19 04 26.3973 | +01 10 36.709 | 4.3 | 144 | 3.76 × 2.32 at −15 | 0.037 | ... |
| G037.43+01.51[b] | 18 54 14.2939 | +04 41 40.571 | ... | ... | 0.84 × 0.31 at −7 | 0.946 | 40.0 |
| J1856+0610 | 18 56 31.8388 | +06 10 16.765 | 1.6 | 21 | 0.86 × 0.33 at −8 | 0.115 | ... |
| J1855+0251 | 18 55 35.4364 | +02 51 19.563 | 1.9 | 169 | 0.87 × 0.33 at −7 | 0.111 | ... |
| J1903+0145 | 19 03 53.0633 | +01 45 26.311 | 3.8 | 140 | 0.89 × 0.35 at −9 | 0.023 | ... |
| J1904+0110 | 19 04 26.3973 | +01 10 36.709 | 4.3 | 144 | 0.87 × 0.34 at −7 | 0.042 | ... |
| G043.79−00.12 | 19 11 53.9868 | +09 35 50.325 | ... | ... | 0.83 × 0.32 at −13 | 269.0 | 43.0 |
| J1905+0952 | 19 05 39.8988 | +09 52 08.408 | 1.6 | −80 | 0.90 × 0.41 at −13 | 0.106 | ... |
| J1913+0932 | 19 13 24.0248 | +09 32 45.374 | 0.4 | 98 | 1.83 × 0.59 at −13 | 0.038 | ... |
| J1907+0907 | 19 07 41.9628 | +09 07 12.404 | 1.1 | −115 | 1.02 × 0.37 at −13 | 0.065 | ... |
| G043.89−00.78 | 19 14 26.3925 | +09 22 36.496 | ... | ... | 0.76 × 0.30 at −11 | 3.458 | 48.6 |
| J1907+0907 | 19 07 41.9634 | +09 07 12.396 | 1.7 | −99 | 2.79 × 1.42 at 15 | 0.061 | ... |
| J1922+0841 | 19 22 18.6336 | +08 41 57.376 | 2.1 | 109 | 1.73 × 0.96 at 60 | 0.007 | ... |
| J1905+0952 | 19 05 39.8989 | +09 52 08.406 | 2.2 | −77 | 1.85 × 0.91 at 64 | 0.095 | ... |
| J1908+1201 | 19 08 40.3206 | +12 01 58.861 | 3.0 | −28 | 1.78 × 0.90 at 64 | 0.063 | ... |
| G045.07+00.13 | 19 13 22.0427 | +10 50 53.336 | ... | ... | 0.86 × 0.32 at −14 | 5.629 | 60.4 |
| J1913+0932 | 19 13 24.0240 | +09 32 45.365 | 1.3 | 180 | 0.90 × 0.33 at −9 | 0.040 | ... |
| J1913+1220 | 19 13 16.2384 | +12 20 39.215 | 1.5 | −1 | 1.05 × 0.34 at −16 | 0.019 | ... |
| G045.45+00.05 | 19 14 21.2658 | +11 09 15.872 | ... | ... | 0.77 × 0.32 at 1 | 8.300 | 51.4 |
| J1908+1201 | 19 08 40.3206 | +12 01 58.860 | 1.6 | −58 | 0.85 × 0.38 at −3 | 0.065 | ... |
| J1905+0952 | 19 05 39.8989 | +09 52 08.407 | 2.4 | −121 | 0.84 × 0.38 at −4 | 0.085 | ... |
| J1925+1227 | 19 25 40.8171 | +12 27 38.086 | 3.1 | 65 | 0.83 × 0.39 at 0 | 0.086 | ... |
| J1913+1307 | 19 13 14.0063 | +13 07 47.330 | 2.0 | −8 | 0.83 × 0.37 at −1 | 0.030 | ... |
| G049.19−00.34 | 19 22 57.7705 | +14 16 09.944 | ... | ... | 0.80 × 0.33 at −3 | 22.30 | 68.6 |
| J1917+1405 | 19 17 18.0639 | +14 05 09.759 | 1.4 | −98 | 0.89 × 0.36 at −5 | 0.051 | ... |
| J1924+1540 | 19 24 39.4559 | +15 40 43.941 | 1.5 | 16 | 0.92 × 0.36 at −5 | 0.483 | ... |
| J1925+1227 | 19 25 40.8171 | +12 27 38.085 | 1.9 | 160 | 0.88 × 0.37 at −6 | 0.118 | ... |
| J1930+1532 | 19 30 52.7670 | +15 32 34.428 | 2.3 | 56 | 0.88 × 0.37 at −2 | 0.371 | ... |

**Notes.** Column 1 gives source names, a and b denote 12 GHz methanol and 22 GHz water masers for of G037.43+01.51; columns 2 and 3 give J2000 coordinates; $\theta_{sep}$ and P.A. in columns 4 and 5 give the magnitude and direction (measured east of north) of the position offset of a background source from the target maser; column 6 gives the full-width at half-maximum of the major and minor axes of the synthesized interferometer beam, with the position angle (east of north) of the major axis listed in degrees; brightnesses given in column 7 are from first-epoch observations; Column 8 gives LSR velocities of the brightest maser spots.

Table 2: Parallaxes and proper motions of high-mass star-forming Regions

| Source | Other | Maser | Parallax (mas) | Distance (kpc) | $\mu_x$ (mas yr$^{-1}$) | $\mu_y$ (mas yr$^{-1}$) | $V_{LSR}$ (km s$^{-1}$) | Ref. |
|---|---|---|---|---|---|---|---|---|
| G351.44+00.65 | NGC6334 | 12 GHz | 0.744 ± 0.076 | $1.35^{+0.05}_{-0.05}$ | 0.31 ± 0.58 | −2.17 ± 0.90 | −8 ± 3 | 1 |
| G011.49−01.48 | | 12 GHz | 0.800 ± 0.033 | $1.25^{+0.05}_{-0.05}$ | 1.42 ± 0.52 | −0.60 ± 0.65 | 11 ± 3 | 1 |
| G014.33−00.64 | | 22 GHz | 0.893 ± 0.102 | $1.12^{+0.14}_{-0.11}$ | 0.95 ± 1.50 | −2.40 ± 1.30 | 22 ± 5 | 2 |
| G014.63−00.57 | | 22 GHz | 0.546 ± 0.022 | $1.83^{+0.08}_{-0.07}$ | 0.22 ± 1.20 | −2.07 ± 1.20 | 19 ± 5 | 1 |
| G015.03−00.67 | M17 | 12 GHz | 0.505 ± 0.033 | $1.98^{+0.14}_{-0.12}$ | 0.68 ± 0.32 | −1.42 ± 0.33 | 22 ± 3 | 3 |
| G034.39+00.22 | | 22 GHz | 0.643 ± 0.049 | $1.56^{+0.13}_{-0.11}$ | −0.90 ± 1.00 | −2.75 ± 2.00 | 57 ± 5 | 4 |
| G035.02+00.34 | | 22 GHz | 0.430 ± 0.040 | $2.32^{+0.23}_{-0.20}$ | −0.92 ± 0.90 | −3.61 ± 0.90 | 52 ± 5 | 1 |
| G035.19−00.74 | | 12 GHz | 0.456 ± 0.045 | $2.19^{+0.24}_{-0.20}$ | −0.18 ± 0.50 | −3.63 ± 0.50 | 30 ± 7 | 5 |
| G035.20−01.73 | W48 | 12 GHz | 0.306 ± 0.045 | $3.27^{+0.56}_{-0.42}$ | −0.71 ± 0.21 | −3.61 ± 0.26 | 42 ± 3 | 5 |
| G037.43+01.51 | | 12 & 22 GHz | 0.532 ± 0.021 | $1.88^{+0.08}_{-0.08}$ | −0.45 ± 0.35 | −3.69 ± 0.39 | 41 ± 3 | 1 |
| G043.79−00.12 | OH43.8-0.1 | 22 GHz | 0.166 ± 0.005 | $6.02^{+0.26}_{-0.24}$ | −3.02 ± 0.36 | −6.20 ± 0.36 | 44 ± 10 | 1 |
| G043.89−00.78 | | 22 GHz | 0.121 ± 0.020 | $8.26^{+1.64}_{-1.17}$ | −2.75 ± 0.30 | −6.43 ± 0.30 | 54 ± 5 | 1 |
| G045.07+00.13 | | 22 GHz | 0.129 ± 0.007 | $7.75^{+0.44}_{-0.40}$ | −3.21 ± 0.26 | −6.11 ± 0.26 | 59 ± 5 | 1 |
| G045.45+00.05 | | 22 GHz | 0.119 ± 0.017 | $8.40^{+1.40}_{-1.05}$ | −2.34 ± 0.38 | −6.00 ± 0.54 | 55 ± 7 | 1 |
| G049.19−00.34 | | 22 GHz | 0.192 ± 0.009 | $5.21^{+0.26}_{-0.23}$ | −2.99 ± 0.40 | −5.71 ± 0.40 | 67 ± 5 | 1 |
| G049.48−00.36 | W51 IRS2 | 12 GHz | 0.195 ± 0.071 | $5.1^{+2.9}_{-1.4}$ | −2.49 ± 0.14 | −5.51 ± 0.16 | 56 ± 3 | 6 |
| G049.48−00.38 | W51 MAIN | 22 GHz | 0.185 ± 0.010 | $5.41^{+0.31}_{-0.28}$ | −2.64 ± 0.20 | −5.11 ± 0.20 | 58 ± 4 | 7 |
| G052.10+01.04 | | 22 GHz | 0.251 ± 0.060 | $3.98^{+0.67}_{-0.50}$ | −2.60 ± 2.00 | −6.10 ± 2.00 | 42 ± 5 | 8 |

**Notes.** Columns 1 and 2 give (Galactic) source names and an alias where appropriate. Column 3 gives the maser transitions: 12 GHz and 22 GHz are $CH_3OH$ and $H_2O$ masers, respectively. Columns 4 and 5 give the trigonometric parallaxes and distances. Columns 6, 7, and 8 give the proper motion components in the eastward ($\mu_x = \mu_\alpha \cos\delta$) and northward directions ($\mu_y = \mu_\delta$) and the LSR velocity of the central exciting star. References are given in Column 9.
References: (1) this paper; (2) Sato et al. 2010a; (3) Xu et al. 2011; (4) Kurayama et al. 2011; (5) Zhang et al. 2009; (6) Xu et al. 2009; (7) Sato et al. 2010b; (8) Oh et al. 2010;

Table 3: Derived parameters of Sagittarius arm sources

| Source | Distance (kpc) | G.C. Radius (kpc) | $\beta$ (deg) | Arm Offset (kpc) |
|---|---|---|---|---|
| G351.44+00.65 | $1.35^{+0.15}_{-0.13}$ | $6.97^{+0.12}_{-0.15}$ | $-1.65^{+0.18}_{-0.23}$ | 0.01 ± 0.09 |
| G011.49−01.48 | $1.25^{+0.05}_{-0.05}$ | $7.08^{+0.05}_{-0.05}$ | $2.02^{+0.10}_{-0.09}$ | 0.17 ± 0.05 |
| G014.33−00.64 | $1.12^{+0.14}_{-0.11}$ | $7.22^{+0.11}_{-0.14}$ | $2.20^{+0.33}_{-0.25}$ | 0.32 ± 0.12 |
| G014.63−00.57 | $1.83^{+0.07}_{-0.06}$ | $6.54^{+0.07}_{-0.07}$ | $4.05^{+0.22}_{-0.20}$ | −0.33 ± 0.07 |
| G015.03−00.67 | $1.98^{+0.13}_{-0.12}$ | $6.55^{+0.11}_{-0.13}$ | $4.60^{+0.42}_{-0.36}$ | −0.45 ± 0.12 |
| G034.39+00.22 | $1.56^{+0.13}_{-0.11}$ | $7.07^{+0.08}_{-0.10}$ | $7.14^{+0.70}_{-0.58}$ | 0.24 ± 0.08 |
| G035.02+00.34 | $2.32^{+0.24}_{-0.20}$ | $6.54^{+0.14}_{-0.16}$ | $11.76^{+1.55}_{-1.23}$ | −0.22 ± 0.13 |
| G035.19−00.74 | $2.19^{+0.24}_{-0.20}$ | $6.63^{+0.14}_{-0.16}$ | $11.00^{+1.54}_{-1.20}$ | −0.14 ± 0.12 |
| G035.20−01.73 | $3.27^{+0.56}_{-0.42}$ | $5.94^{+0.25}_{-0.31}$ | $18.50^{+4.63}_{-3.13}$ | −0.71 ± 0.22 |
| G037.43+01.51 | $1.88^{+0.08}_{-0.08}$ | $6.90^{+0.05}_{-0.05}$ | $9.53^{+0.47}_{-0.43}$ | 0.11 ± 0.04 |
| G043.79−00.12 | $6.02^{+0.19}_{-0.18}$ | $5.74^{+0.01}_{-0.91}$ | $46.53^{+1.86}_{-1.76}$ | −0.50 ± 0.03 |
| G043.89−00.78 | $8.26^{+1.64}_{-1.17}$ | $6.19^{+0.77}_{-0.33}$ | $67.74^{+12.6}_{-10.7}$ | 0.22 ± 0.99 |
| G045.07+00.13 | $7.75^{+0.44}_{-0.40}$ | $6.17^{+0.15}_{-0.11}$ | $62.76^{+3.84}_{-3.59}$ | 0.14 ± 0.15 |
| G045.45+00.05 | $8.40^{+1.40}_{-1.05}$ | $6.45^{+0.68}_{-0.34}$ | $68.12^{+10.3}_{-9.10}$ | 0.49 ± 0.87 |
| G049.19−00.34 | $5.21^{+0.26}_{-0.23}$ | $6.28^{+0.12}_{-0.12}$ | $38.80^{+2.34}_{-2.12}$ | −0.07 ± 0.02 |
| G049.48−00.36 | $5.12^{+2.94}_{-1.37}$ | $6.32^{+0.53}_{-0.20}$ | $39.60^{+25.3}_{-12.1}$ | −0.06 ± 0.80 |
| G049.48−00.38 | $5.40^{+0.31}_{-0.28}$ | $6.31^{+0.01}_{-0.01}$ | $40.63^{+2.80}_{-2.51}$ | −0.03 ± 0.03 |
| G052.10+01.04 | $3.98^{+0.67}_{-0.50}$ | $6.64^{+0.10}_{-0.07}$ | $28.24^{+5.75}_{-4.19}$ | 0.13 ± 0.08 |

**Notes.** Column 2 lists heliocentric distance; Columns 3 and 4 give Galactocentric radius and azimuth; Column 5 is the spatial offset perpendicular to the arm, a positive value locates a source radially outside the spiral arm model. Values given in columns 3, 4, and 5 are calculated assuming $R_0$ = 8.33 kpc.

Table 4: Peculiar motions

| Source | $U_s$ (km s$^{-1}$) | $V_s$ (km s$^{-1}$) | $W_s$ (km s$^{-1}$) |
|---|---|---|---|
| G351.44+00.65 | 0.6 ± 3.6 | 4.5 ± 4.5 | −1.7 ± 4.3 |
| G011.49−01.48 | −0.5 ± 3.6 | 12.8 ± 4.1 | −0.3 ± 3.9 |
| G014.33−00.64 | 11.3 ± 5.6 | 6.3 ± 7.4 | −1.9 ± 8.2 |
| G014.63−00.57 | 2.4 ± 6.0 | −0.2 ± 10.3 | −1.5 ± 10.6 |
| G015.03−00.67 | 0.8 ± 3.9 | 6.7 ± 3.7 | −3.3 ± 3.7 |
| G034.39+00.22 | 22.9 ± 10.2 | 22.2 ± 10.2 | 5.5 ± 9.7 |
| G035.02+00.34 | 11.6 ± 9.2 | 7.0 ± 8.3 | −0.2 ± 10.3 |
| G035.19−00.74 | −4.3 ± 7.4 | −5.1 ± 7.0 | −7.0 ± 5.8 |
| G035.20−01.73 | −11.7 ± 11.5 | -10.2 ± 7.6 | −7.8 ± 4.8 |
| G037.43+01.51 | 7.4 ± 4.0 | 6.5 ± 3.8 | −2.1 ± 3.8 |
| G043.79−00.12 | 19.6 ± 12.3 | -32.6 ± 10.3 | 3.1 ± 10.5 |
| G043.89−00.78 | 9.6 ± 15.0 | −2.5 ± 20.9 | -13.4 ± 12.6 |
| G045.07+00.13 | 8.0 ± 12.4 | −0.1 ± 8.3 | 9.1 ± 9.9 |
| G045.45+00.05 | -12.9 ± 21.1 | −2.1 ± 18.3 | -19.7 ± 17.6 |
| G049.19−00.34 | 9.0 ± 11.7 | 5.7 ± 5.5 | 6.9 ± 10.2 |
| G049.48−00.36 | −0.3 ± 16.8 | −4.1 ± 11.3 | −1.5 ± 5.8 |
| G049.48−00.38 | -10.0 ± 8.4 | -2.3 ± 4.7 | 6.1 ± 5.5 |
| G052.10+01.04 | 16.1 ± 41.0 | -10.7 ± 12.8 | −2.0 ± 42.0 |

**Notes.** $U_s$, $V_s$, $W_s$ give the estimated peculiar (non-circular) velocity components of the central star that excites the masers, assuming $R_0$ = 8.33 ± 0.16 kpc, $\Theta_0$ = 243 ± 6 km s$^{-1}$, and solar motion components ($U_\odot$, $V_\odot$, $W_\odot$) = (10.7 ± 1.8, 12.2 ± 2.0, 8.7 ± 0.9) km s$^{-1}$. $U_s$ is locally toward the Galactic center, $V_s$ is in the direction of Galactic rotation, and $W_s$ is toward the north Galactic pole.

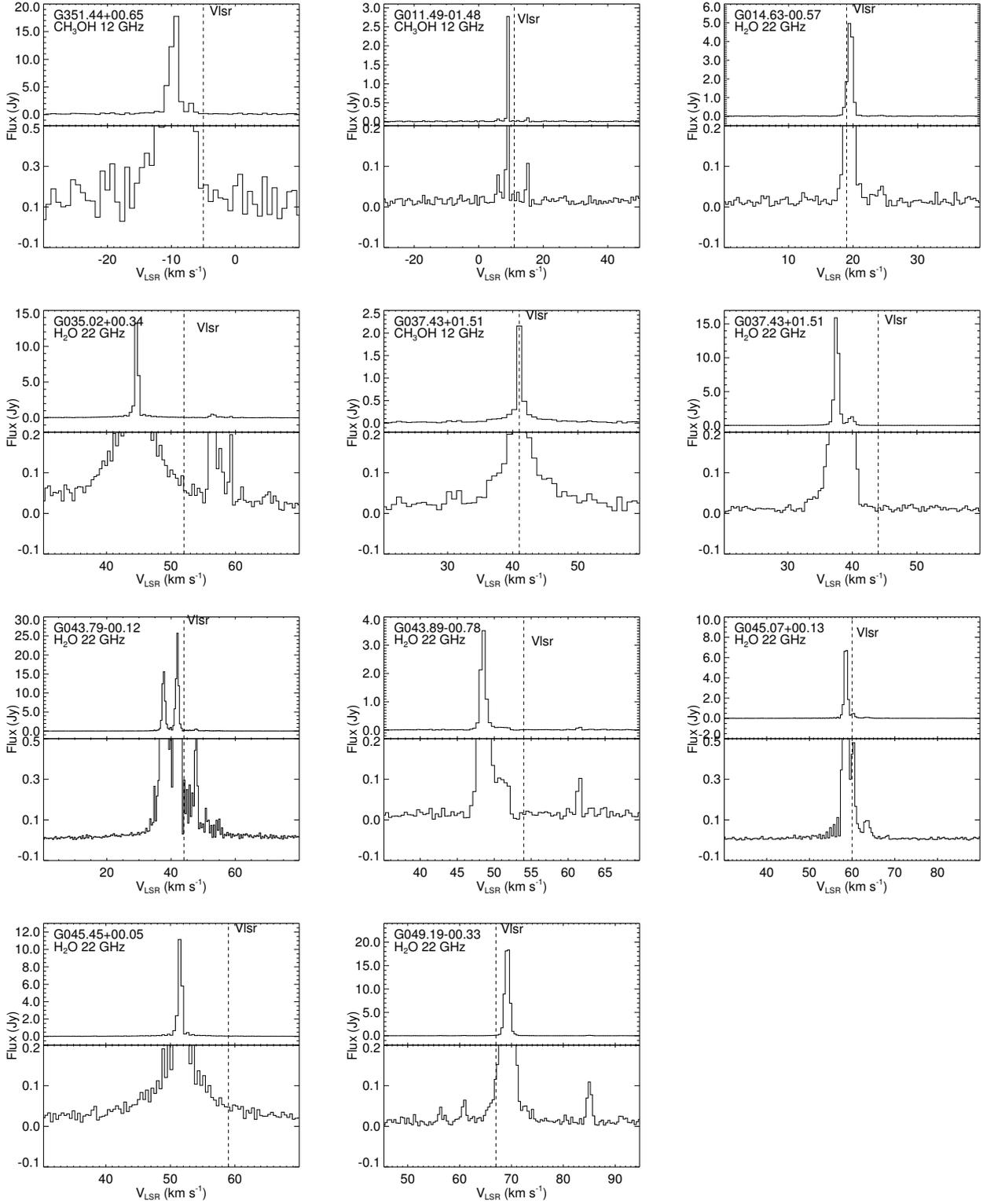

Fig. A.1: Average cross-power (Stokes I) spectra of the eleven observed 12 GHz and 22 GHz masers from the first epoch. Upper panels: full amplitude scale. Lower panels: enlarged view of the weaker spectral features. Dashed lines mark the LSR velocities (from CO line data) of molecular cores hosting the star-forming regions.

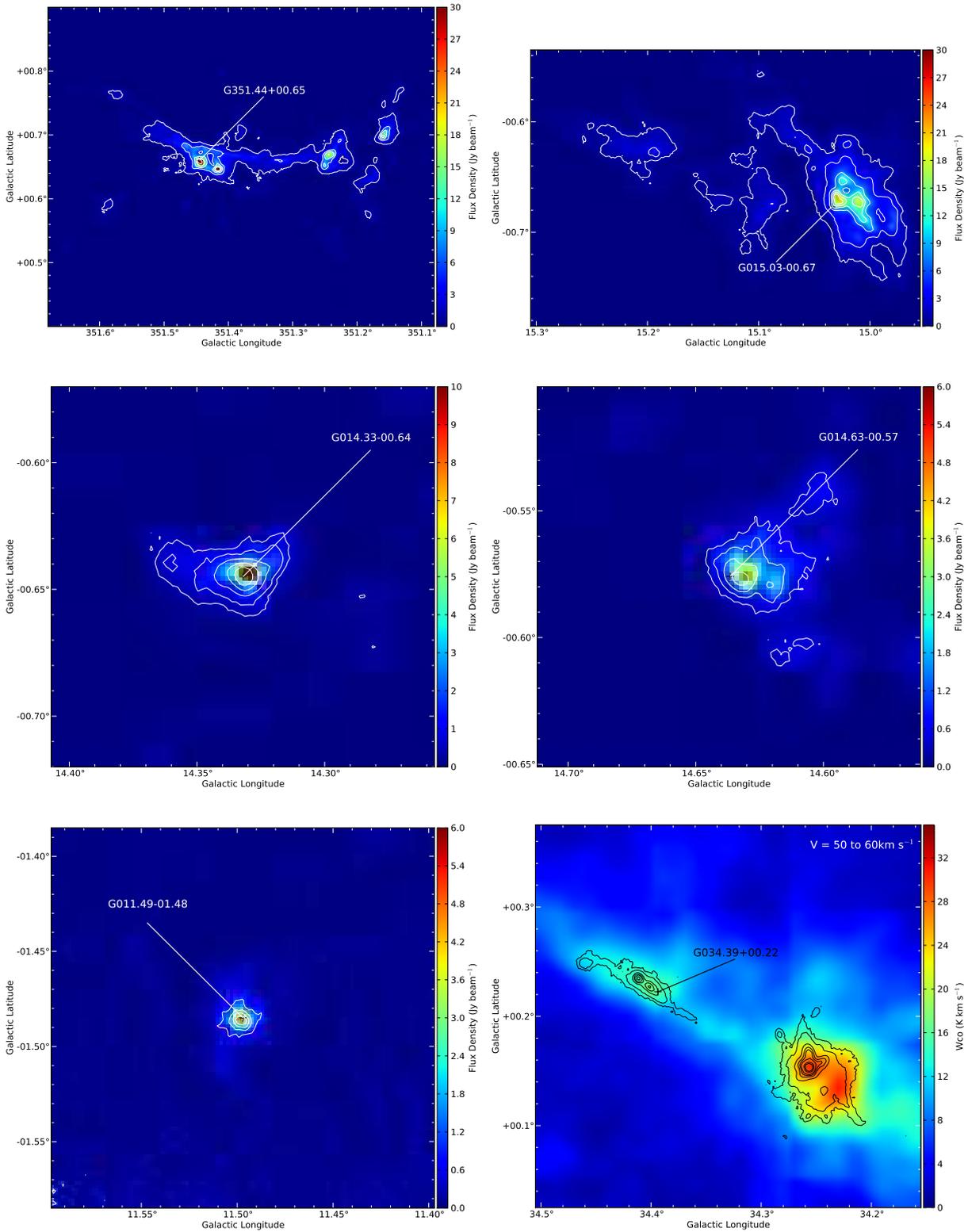

Figure A.2 Maser positions overlay on dust and/or molecular emissions. The position of the maser is indicated with a cross. *Top left panel*: the NGC6334 complex, color image and contours indicate 870 $\mu$m dust emission from the ATLASGAL survey (Schuller et al. 2009), with contour levels of 1, 5, 10, 20, 30 Jy beam$^{-1}$. *Top right panel*: G015.03−00.67 overlaid on 870 $\mu$m dust emission with contour levels of 1, 3, 7,11, 15, 19 Jy beam$^{-1}$. *Top right panel*: G011.49−01.48, color image and contours indicate 870 $\mu$m dust emission, with contour levels of 1, 2, 3, 4 Jy beam$^{-1}$. *Middle left panel*: G014.33−00.64, color image and contours indicate 870 $\mu$m dust emission, with contour levels of 0.5, 1, 2, 4, 8 Jy beam$^{-1}$. *Middle right panel*: G014.63−00.57, color image and contours indicate 870 $\mu$m dust emission, with contour levels of 0.5, 1, 2, 4 Jy beam$^{-1}$. *Bottom left panel*: G015.03−00.67 overlaid on 870 $\mu$m dust emission with contour levels of 1, 3, 7,11, 15, 19 Jy beam$^{-1}$. *Bottom left panel*: G011.49−01.48, color image and contours indicate 870 $\mu$m dust emission, with contour levels of 1, 2, 3, 4 Jy beam$^{-1}$. *Bottom right panel*:G034.39+00.22, false color indicates the intensity of the $^{13}$CO (J = 1−0) line integrated over 50 - 60 km s$^{-1}$, contours are the 870 $\mu$m dust emission, with contour levels of 0.5, 1, 3, 5, 7, 10, 20, 30 Jy beam$^{-1}$. Figure A.2 continued on next page.

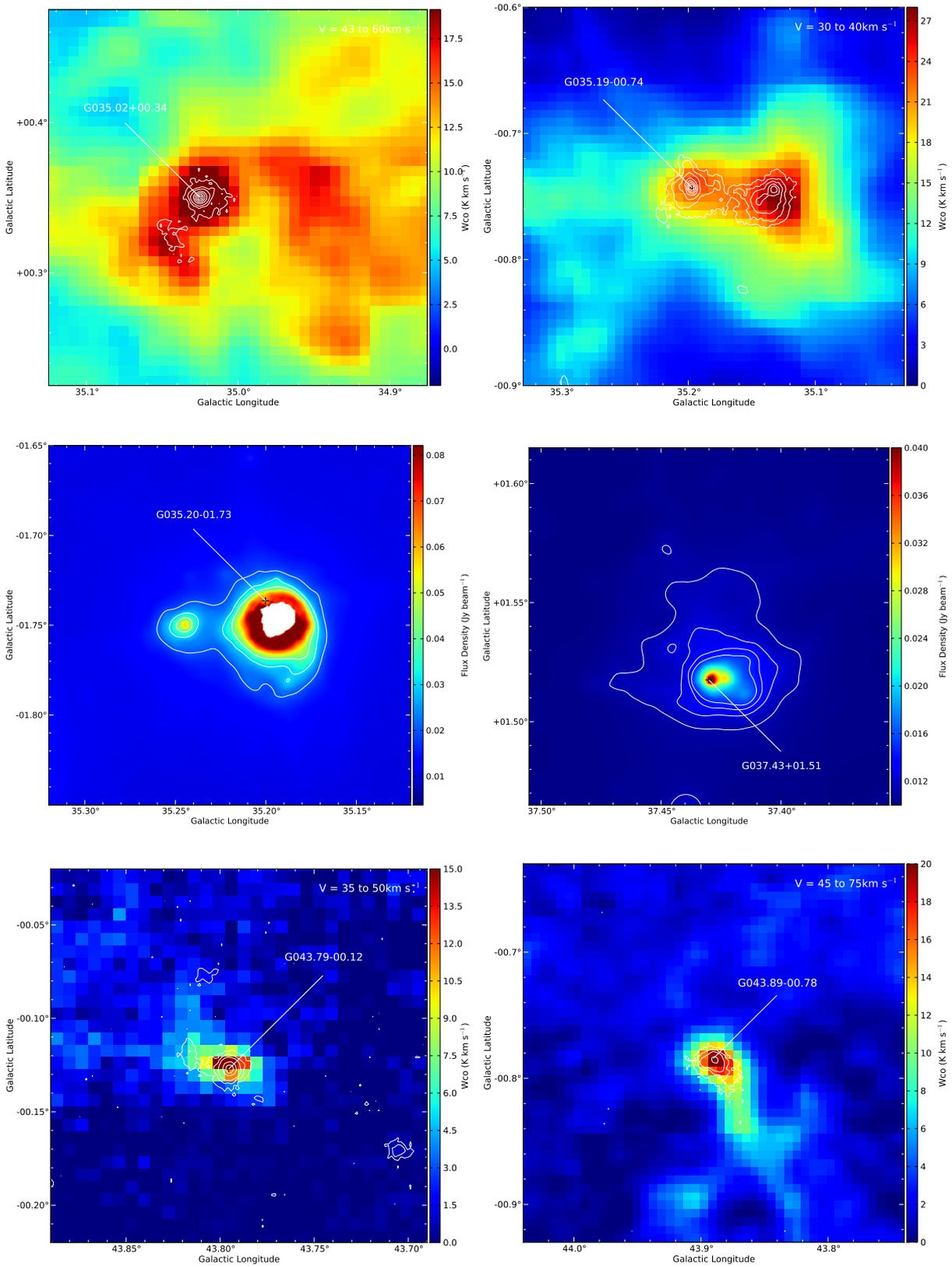

Figure A.2 continued from previous page. *Top left panel*: G035.02+00.34, color image is the intensity of the $^{13}$CO (J = 1−0) emission integrated over $43 < V_{LSR} < 60$ km s$^{-1}$, contours are 870 $\mu$m continuum emission with levels of 0.3, 0.6, 1, 2, 3, 4, 5 Jy beam$^{-1}$. *Top right panel*: G035.19−00.74, color image is the intensity of the $^{13}$CO (J = 1−0) emission integrated over $30 < V_{LSR} < 40$ km s$^{-1}$, contours are 870 $\mu$m dust emission with levels of 0.5, 1, 2, 4, 6, 8, 10 Jy beam$^{-1}$. *Middle left panel:* G035.20−01.73, color image with contours indicate the intensity of 22 $\mu$m dust emission from the Wide-field Infrared Survey Explorer (WISE, Wright et al. 2010), with contour levels of 0.01, 0.02, 0.03 and 0.04 Jy beam$^{-1}$. The central blank pixels are due to saturation. *Middle right panel:* G037.43+01.51, color image is 22 $\mu$m dust emission from WISE, with contour levels of 0.01, 0.02, 0.03 and 0.04 Jy beam$^{-1}$. *Bottom left panel:* G043.79−00.12, color image is $^{13}$CO (J = 1−0) emission integrated over $35 < V_{LSR} < 50$ km s$^{-1}$, contours are 870 $\mu$m dust emission, with levels of 0.2, 0.5, 1, 3, 5 and 7 Jy beam$^{-1}$. *Bottom right panel:* G043.89−00.78, color image is the $^{13}$CO (J = 1−0) emission integrated over t $45 < V_{LSR} < 75$ km s$^{-1}$, contours are 870 $\mu$m dust emission, with levels of 0.3, 0.6, 1, 3, 5 and 7 Jy beam$^{-1}$. Figure A.2 continued on next page.

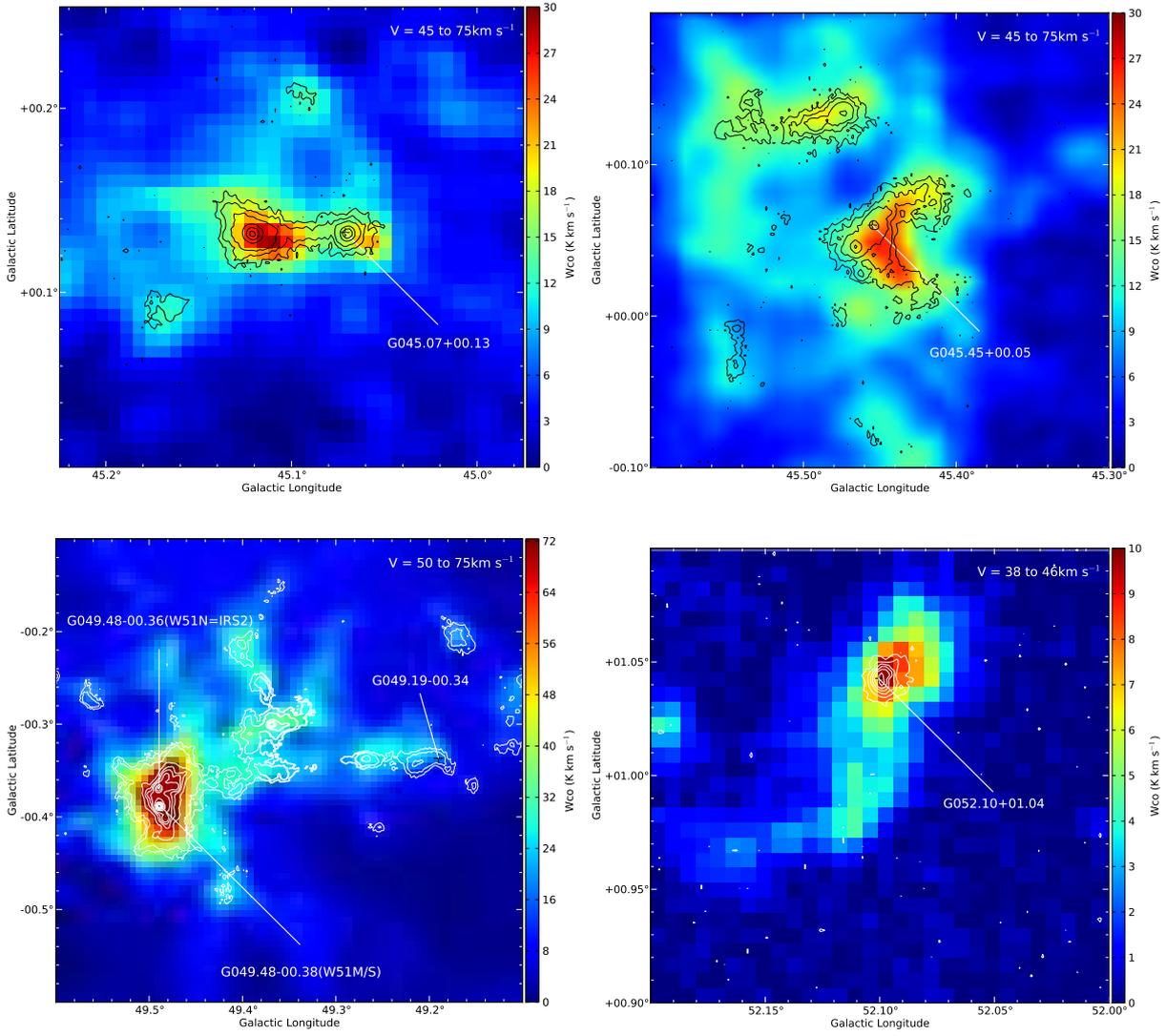

Fig. A.2: Continued from previous page. *Top left panel:* G45.07+00.13, color image indicates the intensity of $^{13}$CO (J = 1−0) emission integrated over 45 < $V_{LSR}$ < 75 km s$^{-1}$, contours are 870 μm continuum emission, with levels of 0.2, 0.5, 1, 3, 5 and 7 Jy beam$^{-1}$. *Top right panel:* G045.45+00.05, color image indicates the intensity of $^{13}$CO (J = 1−0) emission integrated over 45 < $V_{LSR}$ < 75 km s$^{-1}$, contours are 870 μm continuum emission with levels of 0.2, 0.5, 1, 3, 5 and 7 Jy beam$^{-1}$. *Bottom left panel:* W51 star-forming complex, color image indicates the intensity of the $^{13}$CO (J = 1−0) emission integrated over 50 < $V_{LSR}$ < 75 km s$^{-1}$, contours are 870 μm continuum emission with levels of 0.5, 1, 3, 5 7, 10, 20, ..., 70 Jy beam$^{-1}$. *Bottom right panel:* G052.10+01.04, color images indicates the intensity $^{13}$CO (J = 1−0) emission integrated over 38 < $_{LSR}$ < 46 km s$^{-1}$, contours are 870 μm continuum emission with levels of 0.2, 0.4, 0.6, 0.8, 1.0 and 1.2 Jy beam$^{-1}$.

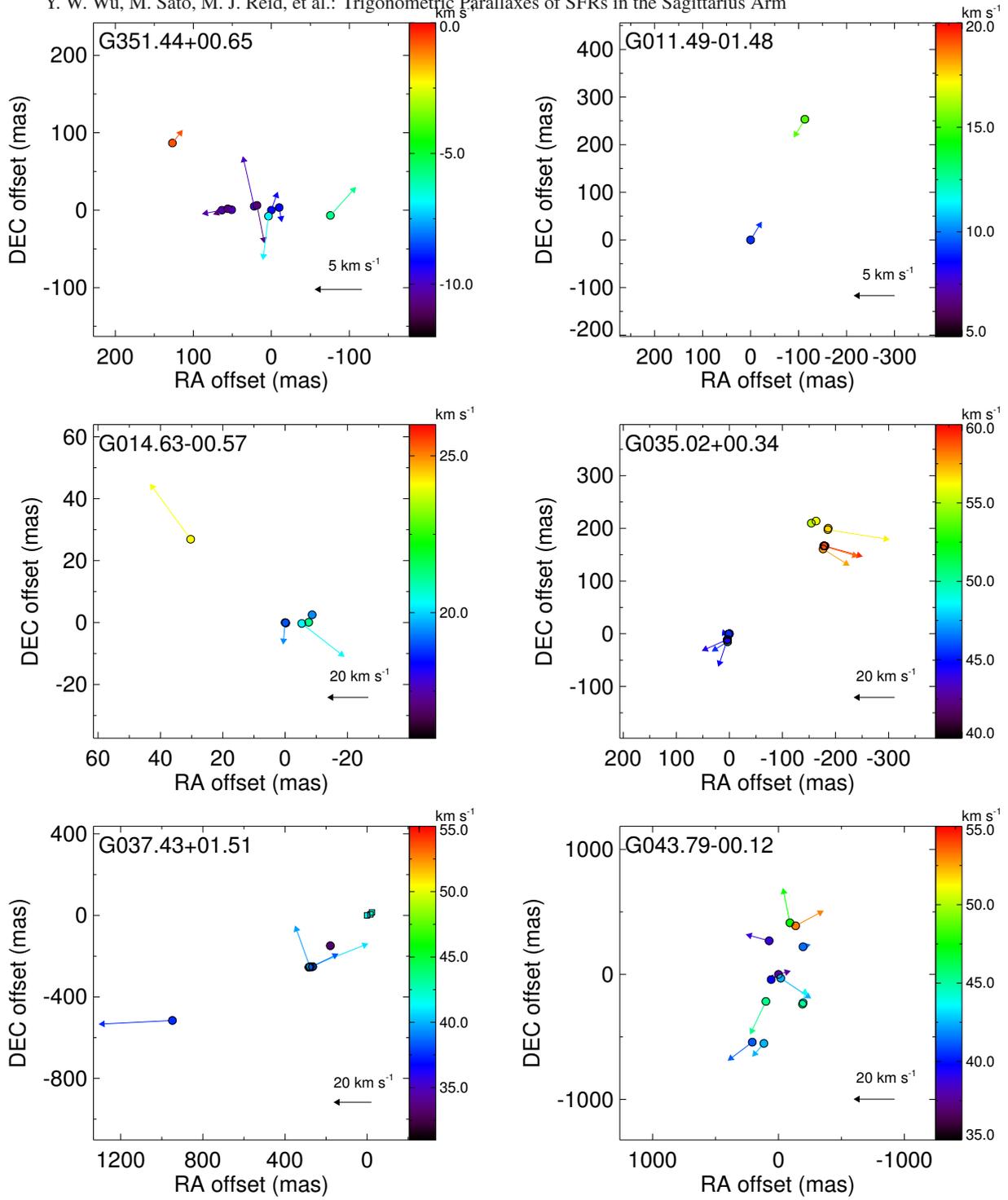

Fig. A.3— Continued

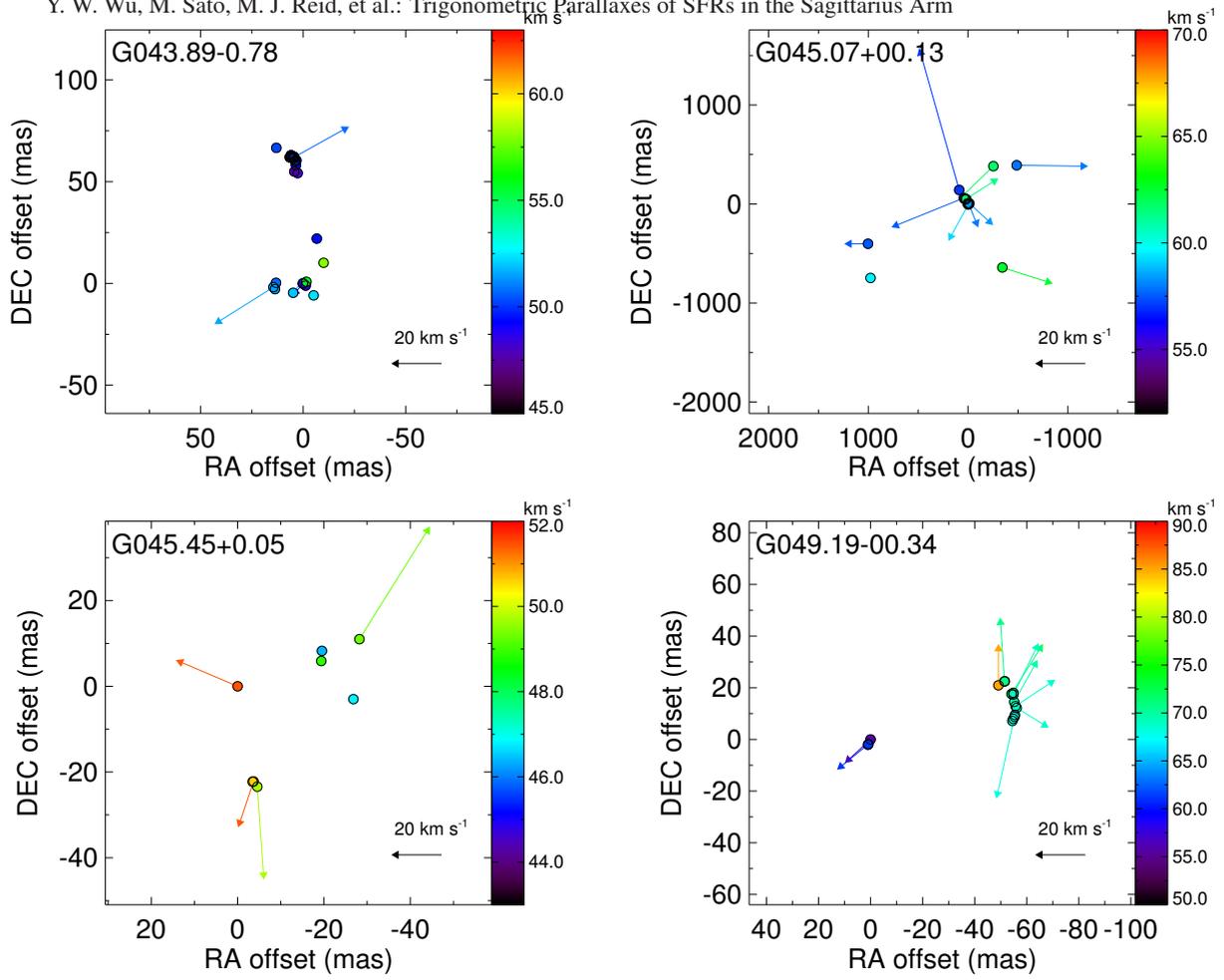

Fig. A.3: Relative proper motions of maser spots, by removing absolute proper motions listed in Table 2. Toward G037.43+1.51, squares and circles denote methanol and water masers, respectively.

Table A.1: Observation information

| Project | Source | Maser | Epoch 1 | Epoch 2 | Epoch 3 | Epoch 4 | Epoch 5 | Epoch 6 | Epoch 7 |
|---|---|---|---|---|---|---|---|---|---|
| BR134A | G043.79−00.12 | $H_2O$ | 2008Oct19 | 2009Apr08 | 2009Apr15 | 2009Aug27 | 2009Oct18 | ... | ... |
| BR145B | G045.07+00.13 | $H_2O$ | 2010Mar13 | 2010Apr03 | 2010Apr30 | 2010Sep05 | 2010Sep12 | 2010Oct03 | 2010Oct17 |
|  | G045.07+00.13 | $H_2O$ | 2010Oct28 | 2010Nov15 | 2011Mar13 | 2011Apr05 | 2011Apr18 | 2011Sep29 | 2011Oct22 |
|  | G045.07+00.13 | $H_2O$ | 2011Nov11 | 2011Nov14 |  |  |  |  |  |
| BR145C | G351.44+00.65 | $CH_3OH$ | 2010Apr09 | 2010Apr13 | 2010Apr17 | 2010Sep11 | 2010Sep25 | 2010Oct29 | 2011Mar11 |
| BR145J | G045.45+00.05 | $H_2O$ | 2011Feb25 | 2011Apr09 | 2011Jun30 | 2011Sep06 | 2011Oct24 | 2011Dec09 | 2012Apr13 |
| BR145L | G035.02+00.34 | $H_2O$ | 2011Apr15 | 2011Jul02 | 2011Oct27 | 2011Dec26 | 2012Apr15 | ... | ... |
| BR145L | G043.89−00.78 | $H_2O$ | 2011Apr15 | 2011Jul02 | 2011Oct27 | 2011Dec26 | 2012Apr15 | ... | ... |
| BR145N | G011.49−01.48 | $CH_3OH$ | 2011Apr01 | 2011Jun27 | 2011Sep02 | 2011Oct17 | 2011Dec08 | 2012Apr08 | ... |
| BR145N | G049.19−00.34 | $H_2O$ | 2011Apr01 | 2011Jun27 | 2011Sep02 | 2011Oct17 | 2011Dec08 | 2012Apr08 | ... |
| BR145T | G014.63−00.57 | $H_2O$ | 2011Mar29 | 2011Jun26 | 2011Aug27 | 2011Oct15 | 2011Nov30 | 2012Apr02 | ... |
| BR145T | G037.43+01.51 | $CH_3OH$ | 2011Mar29 | 2011Jun26 | 2011Aug27 | 2011Oct15 | 2011Nov30 | 2012Apr02 | ... |
| BR145T | G037.43+01.51 | $H_2O$ | 2011Mar29 | 2011Jun26 | 2011Aug27 | 2011Oct15 | 2011Nov30 | 2012Apr02 | ... |

**Notes.** Column 1 gives project codes of observations; column 2 gives source names; column 3 gives maser species, $CH_3OH$ masers are 12 GHz class II masers; $H_2O$ masers are 22 GHz masers; Column 4-10 give observation dates. For G045.07+00.13, we have 16 epochs within 1.5 years.

Table A.2: Detailed results of parallax and proper motion

| Maser | $V_{LSR}$ (km s$^{-1}$) | Background Source | Parallax (mas) | $\mu_x$ (mas yr$^{-1}$) | $\mu_y$ (mas yr$^{-1}$) |
|---|---|---|---|---|---|
| G351.44+0.65 | −9.27 | J1712−3514 | 0.761 ± 0.064 | 0.21 ± 0.17 | −1.87 ± 0.75 |
| | −10.14 | J1712−3514 | 0.730 ± 0.099 | 0.30 ± 0.27 | −2.20 ± 0.75 |
| | −10.18 | J1712−3514 | 0.750 ± 0.081 | 0.61 ± 0.22 | −2.25 ± 0.94 |
| | −10.10 | J1712−3514 | 0.747 ± 0.093 | 0.52 ± 0.25 | −1.34 ± 0.73 |
| | −10.77 | J1712−3514 | 0.719 ± 0.046 | 0.25 ± 0.12 | −2.66 ± 0.67 |
| | −9.21 | J1712−3514 | 0.737 ± 0.086 | 0.29 ± 0.23 | −2.29 ± 0.67 |
| | −0.75 | J1712−3514 | 0.753 ± 0.077 | 0.20 ± 0.21 | −1.80 ± 0.68 |
| | | Combined | **0.744 ± 0.076** | | |
| G011.49−01.48 | 9.08 | J1808−1822 | 0.836 ± 0.036 | 1.03 ± 0.08 | −0.39 ± 0.61 |
| | 9.08 | J1821−2100 | 0.771 ± 0.044 | 1.29 ± 0.11 | 0.08 ± 0.73 |
| | 9.08 | **above 2 comb.** | **0.800 ± 0.033** | 1.16 ± 0.07 | −0.16 ± 0.41 |
| | 9.08 | J1825−1718 | 0.775 ± 0.153 | 1.52 ± 0.37 | −0.91 ± 0.80 |
| | 9.08 | J1832−2039 | 0.949 ± 0.218 | 1.34 ± 0.53 | −0.69 ± 1.22 |
| G014.63−00.57 | 19.00 | J1825−1718 | 0.558 ± 0.032 | 0.52 ± 0.08 | −3.49 ± 0.32 |
| | 19.00 | J1809−1520 | 0.611 ± 0.010 | 0.54 ± 0.01 | −2.98 ± 0.32 |
| | 19.00 | J1810−1626 | 0.572 ± 0.028 | 0.42 ± 0.07 | −3.10 ± 0.22 |
| | 19.42 | J1825−1718 | 0.493 ± 0.042 | 0.30 ± 0.10 | −3.77 ± 0.40 |
| | 19.42 | J1809−1520 | 0.565 ± 0.027 | 0.31 ± 0.07 | −3.21 ± 0.32 |
| | 19.42 | J1810−1626 | 0.509 ± 0.034 | 0.19 ± 0.08 | −3.39 ± 0.32 |
| | 19.84 | J1825−1718 | 0.538 ± 0.033 | 0.34 ± 0.08 | −3.60 ± 0.31 |
| | 19.84 | J1809−1520 | 0.551 ± 0.028 | 0.19 ± 0.07 | −3.36 ± 0.32 |
| | 19.84 | J1810−1626 | 0.521 ± 0.037 | 0.14 ± 0.09 | −3.55 ± 0.37 |
| | 19.42 | **Combined** | **0.546 ± 0.022** | 0.33 ± 0.03 | −3.38 ± 0.09 |
| G035.02+00.34 | 44.58 | J1855+0251 | 0.482 ± 0.048 | 0.04 ± 0.12 | −3.17 ± 0.09 |
| | 59.32 | J1855+0251 | 0.406 ± 0.027 | −1.97 ± 0.09 | −3.66 ± 0.04 |
| | | **Combined** | **0.430 ± 0.040** | | |
| 12 GHz G037.43+01.51 | 41.00 | J1856+0610 | 0.575 ± 0.113 | −0.57 ± 0.25 | −3.62 ± 0.38 |
| | 41.00 | J1855+0251 | 0.430 ± 0.045 | −0.34 ± 0.10 | −3.77 ± 0.37 |
| 22 GHz G037.43+01.51 | 37.00 | J1856+0610 | 0.527 ± 0.007 | 3.96 ± 0.01 | −3.76 ± 0.26 |
| | 37.00 | J1855+0251 | 0.532 ± 0.022 | 3.89 ± 0.05 | −3.74 ± 0.37 |
| | 37.00 | J1903+0145 | 0.485 ± 0.008 | 3.93 ± 0.01 | −4.18 ± 0.26 |
| | 37.00 | J1904+0110 | 0.549 ± 0.035 | 3.88 ± 0.08 | −4.10 ± 0.19 |
| | 37.42 | J1856+0610 | 0.558 ± 0.008 | 3.96 ± 0.02 | −3.69 ± 0.45 |
| | 37.42 | J1855+0251 | 0.564 ± 0.032 | 3.89 ± 0.07 | −3.67 ± 0.37 |

Table A.2 – *Continued from previous page*

| Maser | $V_{LSR}$ (km s$^{-1}$) | Background Source | Parallax (mas) | $\mu_x$ (mas yr$^{-1}$) | $\mu_y$ (mas yr$^{-1}$) |
|---|---|---|---|---|---|
| | 37.42 | J1903+0145 | 0.510± 0.073 | 3.97± 0.17 | −4.11± 0.37 |
| | 37.42 | J1904+0110 | 0.583± 0.046 | 3.88± 0.10 | −4.03± 0.37 |
| | | **Combined** | **0.532 ± 0.021** | **−0.44 ± 0.08** | **−3.69 ± 0.20** |
| G043.79−00.12 | 37.78 | J1905+0952 | 0.162± 0.0014 | −3.23± 0.004 | −6.15± 0.07 |
| | 37.78 | J1907+0907 | 0.171± 0.010 | −3.23± 0.030 | −6.14± 0.07 |
| | 37.78 | **Combined** | **0.166 ± 0.005** | **−3.23 ± 0.013** | **−6.14 ± 0.04** |
| | 37.78 | J1913+0932 | 0.141± 0.021 | −3.18± 0.06 | −6.21± 0.15 |
| G043.89−00.78 | 48.58 | J1922+0841 | 0.105± 0.026 | −2.64± 0.10 | −6.32± 0.04 |
| | 48.58 | J1905+0952 | 0.116± 0.034 | −2.64± 0.07 | −6.16± 0.10 |
| | 48.58 | J1907+0907 | 0.113± 0.048 | −2.64± 0.10 | −6.43± 0.12 |
| | 48.58 | **Combined** | **0.121 ± 0.020** | **−2.64 ± 0.04** | **−6.30 ± 0.06** |
| G045.07+00.13 | 60.00 | J1913+0932 | **0.129± 0.007** | −2.98± 0.01 | −6.49± 0.02 |
| G045.45+00.05 | 51.42 | J1905+0952 | 0.155± 0.038 | −1.74± 0.09 | −5.95± 0.13 |
| | 51.42 | J1908+1201 | 0.123± 0.038 | −1.76± 0.09 | −5.80± 0.15 |
| | 51.42 | J1913+1307 | 0.113± 0.047 | −1.79± 0.11 | −5.69± 0.16 |
| | 51.42 | J1925+1227 | 0.103± 0.039 | −1.85± 0.09 | −5.66± 0.19 |
| | 51.42 | **Combined** | **0.119 ± 0.017** | **−1.79 ± 0.04** | **−5.78 ± 0.07** |
| G049.19−00.34 | 55.94 | J1917+1405 | 0.199 ± 0.007 | −2.66 ± 0.02 | −6.13 ± 0.06 |
| | 55.94 | J1924+1540 | 0.206 ± 0.011 | −2.68 ± 0.03 | −6.05 ± 0.02 |
| | 55.94 | J1925+1227 | 0.227 ± 0.020 | −2.70 ± 0.07 | −6.04 ± 0.03 |
| | 55.94 | J1930+1532 | 0.185 ± 0.017 | −2.68 ± 0.05 | −6.05 ± 0.03 |
| | 56.36 | J1917+1405 | 0.189 ± 0.013 | −2.63 ± 0.03 | −6.16 ± 0.07 |
| | 56.36 | J1924+1540 | 0.191 ± 0.018 | −2.65 ± 0.05 | −6.08 ± 0.04 |
| | 56.36 | J1925+1227 | 0.221 ± 0.015 | −2.67 ± 0.06 | −6.07 ± 0.02 |
| | 56.36 | J1930+1532 | 0.168 ± 0.021 | −2.65 ± 0.07 | −6.08 ± 0.03 |
| | 61.42 | J1917+1405 | 0.178 ± 0.017 | −2.75 ± 0.04 | −5.96 ± 0.11 |
| | 61.42 | J1924+1540 | 0.184 ± 0.022 | −2.77 ± 0.06 | −5.88 ± 0.07 |
| | 61.42 | J1925+1227 | 0.194 ± 0.021 | −2.79 ± 0.07 | −5.88 ± 0.03 |
| | 61.42 | J1930+1532 | 0.149 ± 0.022 | −2.77 ± 0.08 | −5.89 ± 0.03 |
| | 61.84 | J1917+1405 | 0.185 ± 0.011 | −2.73 ± 0.03 | −5.88 ± 0.06 |
| | 61.84 | J1924+1540 | 0.189 ± 0.013 | −2.74 ± 0.03 | −5.81 ± 0.03 |
| | 61.84 | J1925+1227 | 0.202 ± 0.022 | −2.76 ± 0.07 | −5.80 ± 0.03 |
| | 61.84 | J1930+1532 | 0.167 ± 0.018 | −2.74 ± 0.05 | −5.81 ± 0.04 |
| | 69.00 | J1917+1405 | 0.179 ± 0.004 | −3.43 ± 0.01 | −4.82 ± 0.11 |

Table A.2 – *Continued from previous page*

| Maser | $V_{LSR}$ (km s$^{-1}$) | Background Source | Parallax (mas) | $\mu_x$ (mas yr$^{-1}$) | $\mu_y$ (mas yr$^{-1}$) |
|---|---|---|---|---|---|
| | 69.00 | J1924+1540 | 0.189 ± 0.006 | −3.45 ± 0.01 | −4.74 ± 0.10 |
| | 69.00 | J1925+1227 | 0.189 ± 0.028 | −3.47 ± 0.07 | −4.74 ± 0.09 |
| | 69.00 | J1930+1532 | 0.167 ± 0.016 | −3.45 ± 0.04 | −4.74 ± 0.10 |
| | 69.84 | J1917+1405 | 0.194 ± 0.035 | −3.28 ± 0.09 | −4.89 ± 0.14 |
| | 69.84 | J1924+1540 | 0.199 ± 0.039 | −3.29 ± 0.10 | −4.81 ± 0.14 |
| | 69.84 | J1925+1227 | 0.200 ± 0.032 | −3.31 ± 0.04 | −4.81 ± 0.12 |
| | 69.84 | J1930+1532 | 0.170 ± 0.044 | −3.29 ± 0.12 | −4.81 ± 0.13 |
| | | Combined | **0.192 ± 0.009** | | |

Note. Column 1 gives the names of masers; Column 2 gives the $V_{LSR}$ of spots; Column 3 gives names of background sources; Column 4 gives parallaxes; Columns 5 and 6 give proper motions in right ascension and declination.

Table A.3: Best-fitting models for H$_2$O maser kinematics in G049.19−00.34

| Source | x (″) | y (″) | $\triangle v_x$ (km s$^{-1}$) | $\triangle v_y$ (km s$^{-1}$) | $V_{LSR}$ (km s$^{-1}$) | $V_{exp}$ (km s$^{-1}$) |
|---|---|---|---|---|---|---|
| G049.19−00.34 | −0.03 ± 0.02 | 0.00 ± 0.02 | −10 ± 6 | 10 ± 6 | 74 ± 6 | 10.0 ± 6 |

**Notes.** Column 1 gives the maser name; x, y in columns 2 and 3 are position of the central star relative to the reference feature 14; $\triangle v_x$ and $\triangle v_y$ in columns 4 and 5 are the relative proper speed of the central star relative to the reference feature 14 in east and north directions respectively. Column 6 gives the LSR velocity of the central star; Column 7 gives the expanding speed of the outflow.

Table A.4: Relative inner motions

| Source | Feature No. | Detection (epochs) | $V_{LSR}$ (km s$^{-1}$) | $F_{peak}$ (Jy beam$^{-1}$) | $\triangle x$ (mas) | $\triangle y$ (mas) | $\mu_x$ (mas yr$^{-1}$) | $\mu_y$ (mas yr$^{-1}$) |
|---|---|---|---|---|---|---|---|---|
| G351.44+00.65 | 01 | <u>1</u>,2,3,4,5,6,7 | −9.27 | 8.62 | 0.00 ± 0.02 | 0.00 ± 0.04 | 0 | 0 |
| | 02 | <u>1</u>,2,3,4,5,6,7 | −10.14 | 8.50 | 50.84 ± 0.09 | 0.50 ± 0.10 | 0.12 ± 0.10 | −0.38 ± 0.11 |
| | 03 | <u>1</u>,2,3,4,5,6,7 | −10.18 | 3.61 | 63.43 ± 0.08 | 0.02 ± 0.10 | 0.44 ± 0.12 | −0.37 ± 0.14 |
| | 04 | <u>1</u>,2,3,4,5,6,7 | −10.75 | 1.58 | 55.79 ± 0.08 | 1.88 ± 0.10 | 0.35 ± 0.11 | −0.40 ± 0.15 |
| | 05 | <u>1</u>,2,3,4,5,6,7 | −10.00 | 1.54 | 21.96 ± 0.09 | 5.00 ± 0.15 | 0.30 ± 0.12 | 0.53 ± 0.20 |
| | 06 | <u>1</u>,2,3,4,5,6,7 | −10.77 | 1.14 | 18.40 ± 0.08 | 6.26 ± 0.10 | −0.02 ± 0.12 | −0.94 ± 0.16 |
| | 07 | <u>1</u>,2,3,4,5,6,7 | −9.21 | 0.99 | −10.06 ± 0.09 | 3.41 ± 0.14 | 0.06 ± 0.11 | −0.55 ± 0.19 |
| | 08 | <u>1</u>,2,3,4,5,6,7 | −6.92 | 0.64 | 3.65 ± 0.08 | −7.69 ± 0.08 | 0.20 ± 0.11 | −1.04 ± 0.15 |
| | 09 | <u>1</u>,2,3,4,5,6,7 | −0.75 | 0.47 | 126.81 ± 0.08 | 86.70 ± 0.10 | −0.06 ± 0.11 | −0.09 ± 0.17 |
| | 10 | <u>1</u>,2,3,4,5,6,7 | −5.82 | 0.52 | −75.64 ± 0.09 | −6.78 ± 0.11 | −0.32 ± 0.12 | 0.17 ± 0.17 |
| | **average** | | | | | | **0.11 ± 0.47** | **−0.31 ± 0.47** |
| G011.49−01.48 | 01 | 1,2,3,4,<u>5</u>,6 | 9.08 | 2.04 | 0.01 ± 0.01 | 0.04 ± 0.02 | 0 | 0 |
| | 02 | <u>1</u>,2,3,4,5,6 | 15.23 | 1.09 | −113.02 ± 0.02 | 253.32 ± 0.06 | 0.52 ± 0.05 | −0.90 ± 0.10 |
| | **average** | | | | | | **0.26 ± 0.51** | **−0.45 ± 0.51** |
| G014.63−00.57 | 01 | 1,2,3,<u>4</u>,5,6 | 19.42 | 14.45 | 0 ± 0.01 | 0 ± 0.01 | 0 | 0 |
| | 02 | 1,2,5,<u>6</u> | 24.06 | 6.38 | 30.31 ± 0.01 | 26.90 ± 0.02 | 2.18 ± 0.28 | 4.45 ± 0.12 |
| | 03 | 2 | 19.00 | 3.43 | −0.21 ± 0.09 | −0.14 ± 0.06 | ... | ... |
| | 04 | 5 | 19.42 | 1.54 | −8.65 ± 0.01 | 2.51 ± 0.11 | ... | ... |
| | 05 | <u>2</u>,3,4 | 20.26 | 0.81 | −5.33 ± 0.02 | −0.29 ± 0.01 | −2.50 ± 0.00 | −0.58 ± 0.11 |
| | 06 | 6 | 21.11 | 0.26 | −7.55 ± 0.01 | 0.09 ± 0.04 | ... | ... |
| | **average** | | | | | | **−0.11 ± 1.20** | **1.31 ± 1.20** |
| G035.02+0.34 | 01 | 1,2,<u>3</u>,4,5 | 44.58 | 21.88 | 0 ± 0.01 | 0 ± 0.01 | 0 | 0 |
| | 02 | <u>1</u>,2,3,4 | 56.38 | 17.97 | −185.40 ± 0.01 | 197.47 ± 0.01 | −3.02 ± 0.16 | −0.44 ± 0.16 |
| | 03 | 3,4,<u>5</u> | 44.58 | 11.42 | 3.86 ± 0.02 | −11.27 ± 0.04 | 0.13 ± 0.14 | −1.21 ± 0.09 |
| | 04 | 1,<u>2</u>,3,4,5 | 45.42 | 5.95 | 3.33 ± 0.01 | −14.95 ± 0.12 | 0.43 ± 0.02 | −0.44 ± 0.02 |
| | 05 | 2 | 49.21 | 5.95 | 3.39 ± 0.01 | −14.98 ± 0.01 | ... | ... |
| | 06 | 1,<u>2</u>,3,4,5 | 59.33 | 4.90 | −178.21 ± 0.03 | 167.02 ± 0.01 | −1.99 ± 0.08 | −0.48 ± 0.06 |
| | 07 | 1,2,<u>3</u>,4,5 | 44.16 | 4.85 | 0.06 ± 0.01 | 0.21 ± 0.01 | 0.07 ± 0.04 | 0.06 ± 0.08 |
| | 08 | 3,4,<u>5</u> | 45.00 | 4.61 | 3.95 ± 0.01 | −12.08 ± 0.03 | −0.07 ± 0.07 | 0.49 ± 0.05 |
| | 09 | <u>1</u>,2,3 | 58.48 | 3.70 | −177.45 ± 0.01 | 166.50 ± 0.00 | −1.81 ± 0.01 | −0.47 ± 0.04 |
| | 10 | <u>1</u>,2,3,4 | 57.64 | 3.57 | −176.47 ± 0.03 | 160.85 ± 0.04 | −1.45 ± 0.03 | −0.75 ± 0.02 |
| | 11 | 4,<u>5</u> | 55.11 | 3.43 | −154.30 ± 0.04 | 209.82 ± 0.01 | ... | ... |
| | 12 | 3 | 58.90 | 2.66 | −178.76 ± 0.01 | 166.51 ± 0.01 | ... | ... |
| | 13 | 2,<u>3</u>,4 | 44.16 | 1.95 | 3.43 ± 0.03 | −10.74 ± 0.03 | 0.87 ± 0.06 | −0.51 ± 0.03 |

Table A.4 – *Continued on next page*

Table A.4 – *Continued from previous page*

| Source | Feature No. | Detection (epochs) | $V_{LSR}$ (km s$^{-1}$) | $F_{peak}$ (Jy beam$^{-1}$) | $\Delta x$ (mas) | $\Delta y$ (mas) | $V_x$ (mas yr$^{-1}$) | $V_y$ (mas yr$^{-1}$) |
|---|---|---|---|---|---|---|---|---|
| | 14 | 5 | 45.42 | 0.97 | 2.11 ± 0.07 | −0.55± 0.06 | ... | ... |
| | 15 | 4 | 56.80 | 0.82 | −186.06± 0.03 | 200.29 ± 0.03 | ... | ... |
| | 16 | 4,5 | 55.96 | 0.54 | −163.25± 0.02 | 213.89 ± 0.01 | ... | ... |
| | 17 | 5 | 58.90 | 0.23 | −180.20± 0.04 | 166.51 ± 0.04 | ... | ... |
| | **average** | | | | | | **−0.91 ± 0.90** | **−0.40 ± 0.90** |
| G037.43+01.51$^a$ | 01 | 1,2,3,4,5,6 | 41.00 | 2.53 | 0.00 ± 0.02 | 0.00 ± 0.03 | 0 | 0 |
| | 02 | 1,3,4,5,6 | 41.77 | 1.07 | −16.27± 0.09 | 5.23 ± 0.14 | ... | ... |
| | 03 | 2,3,4 | 41.77 | 0.98 | −23.50± 0.10 | 14.46 ± 0.12 | ... | ... |
| G037.43+01.51$^b$ | 01 | 1,2,3,4 | 39.54 | 108.52 | −669.62± 0.04 | 262.90 ± 0.06 | −3.49 ± 0.17 | 2.69 ± 0.12 |
| | 02 | 1,2,3,4,5 | 37.43 | 18.93 | 0.00 ± 0.01 | 0.00 ± 0.01 | 0 | 0 |
| | 03 | 3,5 | 38.28 | 11.83 | −665.59± 0.02 | 261.70 ± 0.03 | ... | ... |
| | 04 | 1,2 | 33.22 | 5.38 | −769.34± 0.00 | 366.74 ± 0.00 | ... | ... |
| | 05 | 4 | 33.64 | 3.40 | −665.44± 0.03 | 260.98 ± 0.01 | ... | ... |
| | 06 | 3,4,5 | 40.80 | 3.16 | −672.25± 0.28 | 264.68 ± 0.08 | −7.80 ± 0.81 | 1.59 ± 0.04 |
| | 07 | 2,3,4,5 | 38.28 | 1.77 | −680.78± 0.00 | 263.86 ± 0.01 | −5.93 ± 0.74 | 1.00 ± 0.15 |
| | 08 | 5 | 38.28 | 0.28 | −683.54± 0.02 | 264.24 ± 0.03 | ... | ... |
| | **average** | | | | | | **−4.30 ± 0.34** | **1.32 ± 0.34** |
| G043.79−00.12 | 01 | 1,2,3,4 | 37.7 | 269.2 | 0.0 ± 0.10 | 0.0 ± 0.10 | 0 | 0 |
| | 02 | 1,2 | 38.7 | 96.1 | 74.0 ± 0.20 | 279.0 ± 0.10 | 0.61 ± 0.06 | 0.05 ± 0.01 |
| | 03 | 1,2 | 39.7 | 26.5 | 58.0 ± 0.30 | −41.0± 0.30 | 0.18 ± 0.03 | 0.03 ± 0.02 |
| | 04 | 1,2 | 41.1 | 26.6 | 208.0 ± 0.50 | −542.0± 0.10 | 0.63 ± 0.11 | −0.38 ± 0.02 |
| | 05 | 1,2,3 | 41.4 | 82.7 | −195.0± 0.20 | 221.0 ± 0.10 | 0.08 ± 0.04 | −0.02 ± 0.02 |
| | 06 | 1,2 | 42.3 | 29.6 | −18.0± 0.20 | −30.0± 0.10 | −0.31 ± 0.06 | −0.41 ± 0.01 |
| | 07 | 1,2 | 42.5 | 11.7 | 116.0 ± 0.20 | −552.0± 0.10 | 0.41 ± 0.05 | −0.29 ± 0.03 |
| | 08 | 1,2,3,4,5 | 47.5 | 10.3 | −91.0± 0.10 | 413.0 ± 0.10 | 0.33 ± 0.02 | 0.54 ± 0.01 |
| | 09 | 1,2 | 53.1 | 6.8 | −136.0± 0.10 | 389.0 ± 0.10 | −0.27 ± 0.07 | 0.20 ± 0.02 |
| | 10 | 1,2 | 45.3 | 5.1 | 100.0 ± 0.10 | −226.0± 0.10 | 0.48 ± 0.01 | −0.63 ± 0.02 |
| | 11 | 1,2 | 43.5 | 2.5 | −196.0± 0.20 | −237.0± 0.20 | 0.17 ± 0.06 | 0.19 ± 0.04 |
| | 12 | 1,2 | 45.1 | 1.7 | −191.0± 0.20 | −248.0± 0.10 | 0.27 ± 0.05 | 0.02 ± 0.03 |
| | **average** | | | | | | **0.21 ± 0.35** | **−0.06 ± 0.35** |
| G043.89−00.78 | 01 | 2,3,4,5 | 50.69 | 10.46 | 4.94 ±0.00 | 61.73 ± 0.01 | −0.68 ±0.09 | 0.45 ±0.06 |
| | 02 | 2 | 49.84 | 7.35 | 3.86 ±0.00 | 60.63 ± 0.01 | ... | ... |
| | 03 | 1,2,3,4,5 | 48.58 | 6.87 | 0.00 ±0.00 | 0.00 ± 0.00 | 0 | 0 |
| | 04 | 5 | 50.69 | 1.79 | 4.28 ±0.08 | 61.55 ± 0.44 | ... | ... |

Table A.4 – *Continued from previous page*

| Source | Feature No. | Detection (epochs) | $V_{LSR}$ (km s$^{-1}$) | $F_{peak}$ (Jy beam$^{-1}$) | $\triangle x$ (mas) | $\triangle y$ (mas) | $V_x$ (mas yr$^{-1}$) | $V_y$ (mas yr$^{-1}$) |
|---|---|---|---|---|---|---|---|---|
| | 05 | 4,5 | 50.26 | 1.45 | 5.90 ±0.00 | 62.95 ± 0.02 | ... | ... |
| | 06 | 4,5 | 50.26 | 1.45 | 5.90 ±0.00 | 62.95 ± 0.02 | ... | ... |
| | 07 | 5 | 50.69 | 1.41 | 4.40 ±0.00 | 62.26 ± 0.00 | ... | ... |
| | 08 | 1,2 | 61.64 | 1.19 | 6.59 ±0.01 | 62.05 ± 0.01 | ... | ... |
| | 09 | 1,2 | 49.42 | 1.15 | −6.67 ±0.01 | 22.02 ± 0.01 | ... | ... |
| | 10 | 1,2 | 49.42 | 1.09 | 3.54 ±0.00 | 58.09 ± 0.01 | ... | ... |
| | 11 | 1,2 | 49.84 | 1.03 | 6.23 ±0.01 | 61.95 ± 0.01 | ... | ... |
| | 12 | 1,2 | 49.84 | 1.03 | 6.23 ±0.01 | 61.95 ± 0.01 | ... | ... |
| | 13 | 1,2 | 50.26 | 1.01 | 13.07 ±0.00 | 66.63 ± 0.00 | ... | ... |
| | 14 | 4 | 48.16 | 0.98 | 0.14 ±0.00 | −0.08± 0.00 | ... | ... |
| | 15 | 5 | 48.16 | 0.93 | 2.67 ±0.01 | 54.22 ± 0.01 | ... | ... |
| | 16 | 5 | 48.58 | 0.75 | −1.23 ±0.13 | −1.07± 0.07 | ... | ... |
| | 17 | 4 | 50.26 | 0.70 | 3.27 ±0.01 | 60.46 ± 0.02 | ... | ... |
| | 18 | 1,2 | 50.69 | 0.66 | 13.23 ±0.01 | 0.29 ± 0.01 | ... | ... |
| | 19 | 1,2,3,4 | 51.53 | 0.62 | 14.50 ±0.01 | −1.83± 0.04 | 0.49 ±0.04 | −0.25 ±0.06 |
| | 20 | 5 | 50.69 | 0.59 | 5.54 ±0.02 | 62.20 ± 0.06 | ... | ... |
| | 21 | 5 | 50.26 | 0.57 | 5.04 ±0.01 | 62.49 ± 0.02 | ... | ... |
| | 22 | 1,2,3 | 51.95 | 0.50 | 4.86 ±0.00 | −4.61± 0.02 | −0.24 ±0.02 | 0.30 ±0.01 |
| | 23 | 3 | 57.85 | 0.27 | −9.98 ±0.02 | 10.13 ± 0.06 | ... | ... |
| | 24 | 4,5 | 47.74 | 0.27 | 4.28 ±0.00 | 54.97 ± 0.02 | ... | ... |
| | 25 | 5 | 51.53 | 0.25 | 13.75 ±0.01 | −2.80± 0.01 | ... | ... |
| | 26 | 1 | 49.84 | 0.17 | 3.79 ±0.01 | 59.92 ± 0.02 | ... | ... |
| | 27 | 3,4 | 52.37 | 0.14 | −5.11 ±0.01 | −5.88± 0.02 | ... | ... |
| | 28 | 4 | 55.74 | 0.10 | −1.60 ±0.01 | 0.80 ± 0.00 | ... | ... |
| | average | | | | | | **−0.11 ± 0.30** | **−0.13 ± 0.30** |
| G045.07+00.13 | 01 | 4,5,6,7,8,9 | 58.31 | 24.31 | −5.66 ± 0.01 | 2.29 ± 0.01 | ... | ... |
| | 02 | 1,...,14,15,16 | 60.42 | 10.75 | 0.00 ± 0.01 | 0.00 ± 0.01 | 0 | 0 |
| | 03 | 1,2,3 | 58.31 | 5.84 | −5.19 ± 0.01 | 2.24 ± 0.01 | ... | ... |
| | 04 | 4,5,6,7,8,9 | 57.89 | 5.19 | −5.69 ± 0.01 | 2.30 ± 0.01 | −0.33 ± 0.04 | 0.13 ± 0.01 |
| | 05 | 4,5,6,7,8,9 | 57.89 | 3.13 | −486.70 ± 0.01 | 391.39 ± 0.01 | −0.98 ± 0.08 | 0.37 ± 0.06 |
| | 06 | 1,2,3,4,5,...,9 | 58.31 | 2.63 | −12.33± 0.01 | 6.00 ± 0.01 | −0.48 ± 0.01 | 0.15 ± 0.01 |
| | 07 | 10,11,12 | 58.31 | 2.31 | −5.42 ± 0.01 | 2.35 ± 0.01 | ... | ... |
| | 08 | 10,14,15,16 | 59.58 | 1.79 | 977.98 ± 0.02 | −746.88 ± 0.03 | ... | ... |
| | 09 | 4,5,6,7,8,9 | 61.26 | 1.78 | 21.01 ± 0.01 | 49.34 ± 0.01 | −0.57 ± 0.01 | 0.60 ± 0.03 |

Table A.4 – *Continued from previous page*

| Source | Feature No. | Detection (epochs) | $V_{LSR}$ (km s$^{-1}$) | $F_{peak}$ (Jy beam$^{-1}$) | $\triangle x$ (mas) | $\triangle y$ (mas) | $V_x$ (mas yr$^{-1}$) | $V_y$ (mas yr$^{-1}$) |
|---|---|---|---|---|---|---|---|---|
| | 10 | 1,2,<u>3</u> | 57.89 | 1.11 | −5.71 ± 0.01 | 2.27 ± 0.01 | ... | ... |
| | 11 | 2,3,<u>4</u>,5 | 59.16 | 1.11 | −2.94 ± 0.01 | −0.38 ± 0.02 | −0.02 ± 0.02 | −0.01 ± 0.01 |
| | 12 | 14,<u>15</u>,16 | 57.47 | 0.99 | 42.76 ± 0.01 | 60.56 ± 0.01 | 0.53 ± 0.01 | 0.07 ± 0.01 |
| | 13 | 4,5,<u>6</u> | 60.42 | 0.96 | 0.45 ± 0.01 | 0.06 ± 0.01 | ... | ... |
| | 14 | 2,<u>3</u> | 55.79 | 0.87 | 36.25 ± 0.01 | 50.53 ± 0.01 | ... | ... |
| | 15 | 4,5,6,<u>7</u>,8 | 57.05 | 0.75 | 89.15 ± 0.01 | 141.75 ± 0.01 | 0.19 ± 0.02 | 1.88 ± 0.05 |
| | 16 | <u>2</u>,3,4,5,6,7,8 | 62.53 | 0.68 | −343.08 ± 0.01 | −640.96 ± 0.01 | −0.76 ± 0.01 | 0.21 ± 0.02 |
| | 17 | 11,12,<u>13</u>,...,19 | 61.69 | 0.58 | −253.01 ± 0.02 | 382.04 ± 0.02 | 0.12 ± 0.02 | 0.04 ± 0.01 |
| | 18 | 9,14,15,16 | 57.47 | 0.49 | 1003.30 ± 0.01 | −401.46 ± 0.01 | 0.02 ± 0.02 | 0.38 ± 0.02 |
| | 19 | 6,8 | 60.42 | 0.39 | −1.17 ± 0.02 | −6.17 ± 0.05 | ... | ... |
| | 20 | <u>10</u>,11,12 | 57.89 | 0.37 | −5.49 ± 0.01 | 2.37 ± 0.01 | ... | ... |
| | 21 | 14,<u>15</u> | 61.69 | 0.33 | 20.64 ± 0.01 | 50.46 ± 0.02 | ... | ... |
| | 22 | 1,<u>2</u> | 59.58 | 0.32 | −0.85 ± 0.01 | 0.24 ± 0.02 | ... | ... |
| | **average** | | | | | | **−0.23 ± 0.18** | **0.38 ± 0.18** |
| G045.45+00.05 | 01 | 1,2,<u>3</u>,4,5,6,7 | 51.42 | 14.25 | 0.00 ± 0.01 | 0.00 ± 0.01 | 0 | 0 |
| | 02 | 4,<u>5</u>,6,7 | 51.42 | 1.77 | −3.70 ± 0.01 | −22.19± 0.02 | −0.41 ± 0.04 | −0.64 ± 0.10 |
| | 03 | <u>1</u>,2,3 | 50.58 | 3.47 | −3.47 ± 0.01 | −22.26± 0.01 | ... | ... |
| | 04 | 5,<u>6</u>,7 | 49.74 | 0.37 | −4.57 ± 0.01 | −23.42± 0.03 | −0.61 ± 0.01 | −1.07 ± 0.07 |
| | 05 | 1 | 48.89 | 0.80 | −19.38 ± 0.11 | 5.90 ± 0.02 | ... | ... |
| | 06 | <u>3</u>,4,5,6,7 | 49.31 | 0.69 | −28.21 ± 0.02 | 10.99 ± 0.02 | −1.19 ± 0.23 | 0.81 ± 0.12 |
| | 07 | 6 | 46.37 | 0.33 | −19.53 ± 0.00 | 8.26 ± 0.01 | ... | ... |
| | 08 | 6 | 46.79 | 0.26 | −26.82 ± 0.01 | −3.00 ± 0.01 | ... | ... |
| | **average** | | | | | | **−0.55 ± 0.38** | **−0.22 ± 0.54** |
| G049.19−00.34 | 01 | 2,<u>3</u>,4,5,6 | 69.00 | 62.14 | −54.86 ± 0.00 | 17.39 ± 0.00 | −0.80 ± 0.01 | 1.22 ± 0.05 |
| | 02 | 4,<u>5</u>,6 | 70.68 | 57.58 | −54.85 ± 0.02 | 18.06 ± 0.01 | ... | ... |
| | 03 | 1,<u>2</u>,3,4 | 69.42 | 28.04 | −55.18 ± 0.00 | 14.54 ± 0.02 | −0.77 ± 0.02 | 1.07 ± 0.04 |
| | 04 | 3,4,5,<u>6</u> | 68.58 | 27.63 | −56.13 ± 0.00 | 12.22 ± 0.10 | −0.91 ± 0.01 | 0.08 ± 0.12 |
| | 05 | 1,<u>2</u>,3,4,5 | 68.58 | 23.43 | −55.68 ± 0.01 | 12.91 ± 0.06 | −1.03 ± 0.02 | 0.82 ± 0.09 |
| | 06 | 1 | 68.16 | 20.09 | −55.44 ± 0.01 | 9.51 ± 0.04 | ... | ... |
| | 07 | 1,2,3,<u>4</u>,5,6 | 71.53 | 16.20 | −54.14 ± 0.02 | 17.59 ± 0.04 | −0.90 ± 0.09 | 1.20 ± 0.07 |
| | 08 | 2,3,4,<u>5</u>,6 | 68.16 | 12.42 | −55.31 ± 0.04 | 8.78 ± 0.08 | −0.11 ± 0.04 | −0.91 ± 0.03 |
| | 09 | 1,2,3,4,<u>5</u>,6 | 70.68 | 8.79 | −51.44 ± 0.01 | 22.45 ± 0.02 | −0.33 ± 0.09 | 1.42 ± 0.08 |
| | 10 | 1,<u>2</u>,3,4,5 | 67.74 | 7.00 | −54.80 ± 0.00 | 8.00 ± 0.02 | ... | ... |
| | 11 | 3,<u>5</u>,6 | 68.16 | 4.02 | −54.47 ± 0.01 | 7.17 ± 0.03 | ... | ... |

Table A.4 – *Continued from previous page*

| Source | Feature No. | Detection (epochs) | $V_{LSR}$ (km s$^{-1}$) | $F_{peak}$ (Jy beam$^{-1}$) | $\triangle x$ (mas) | $\triangle y$ (mas) | $V_x$ (mas yr$^{-1}$) | $V_y$ (mas yr$^{-1}$) |
|---|---|---|---|---|---|---|---|---|
| | 12 | 1,2,3,4,5,6 | 71.11 | 3.28 | −51.53 ± 0.05 | 22.54 ± 0.01 | ... | ... |
| | 13 | 1,2,3,4 | 85.01 | 3.00 | −49.11 ± 0.02 | 20.95 ± 0.02 | −0.40 ± 0.04 | 1.06 ± 0.00 |
| | 14 | 1,2,3,4,5,6 | 55.94 | 1.22 | 0.00 ± 0.01 | 0.00 ± 0.01 | 0 | 0 |
| | 15 | 1,2,3,4,5,6 | 61.00 | 0.57 | 0.97 ± 0.01 | −2.03± 0.02 | ... | ... |
| | 16 | 1,2,3,4,5 | 60.99 | 0.48 | 0.83 ± 0.01 | −1.82± 0.01 | 0.10 ± 0.02 | −0.01 ± 0.03 |
| **average** | | | | | | | **−0.40 ± 0.40** | **0.40 ± 0.40** |

**N**ote. Column 1 gives the source name. Column 2 gives the feature label, increasing with decreasing brightness. Column 3 reports the detected epochs. Columns 4 and 5 list the LSR velocity and brightness of the brightest feature's spot, observed at the epoch that is underlined in Column 3. Columns 6 and 7 give the position offset relative to the reference feature along the east and north directions, respectively, at the first detected epoch. Columns 8 and 9 report the proper motion components relative to the reference feature along the east and north directions, respectively. a and b denote 12 GHz methanol and 22 GHz water masers in field of G037.43+01.51, respectively.